\newcommand{\compage}{
\onecolumn
\widetext 
}

\newcommand{\rgtlne}{
\noindent{\setlength{\unitlength}{1.0pc}
\begin{picture}(20.5,.5)
\thinlines
\put(0,.5){\line(1,0){20.5}}
\put(0,0.5){\line(0,-1){0.5}}
\end{picture}
}}
\newcommand{\lftlne}{
\noindent{\setlength{\unitlength}{1.0pc}
\begin{picture}(20.5,.5)
\thinlines
\put(0,0){\line(1,0){20.5}}
\put(20.5,0){\line(0,1){0.5}}
\end{picture}
}}

\newcommand{\tworef}[2]{
\hspace{-1em}\begin{minipage}[t]{20.00pc}
\marginparpush 5pt \parskip 0pt plus 1pt \parindent 1em \partopsep 2pt 
plus 1pt minus 1pt
#1

\end{minipage}\hspace{1.6pc}
\begin{minipage}[t]{20.00pc}
\marginparpush 5pt \parskip 0pt plus 1pt \parindent 1em \partopsep 2pt 
plus 1pt minus 1pt
#2
\end{minipage}
}
\newcommand{\twomini}[2]{
\hspace{-1em}\begin{minipage}[t]{20.5pc}
\marginparpush 5pt \parskip 0pt plus 1pt \parindent 1em \partopsep 2pt 
plus 1pt minus 1pt
#1

\end{minipage}\hspace{1.3pc}
\begin{minipage}[t]{20.5pc}
\marginparpush 5pt \parskip 0pt plus 1pt \parindent 1em \partopsep 2pt 
plus 1pt minus 1pt
#2
\end{minipage}
}
\documentstyle[prb,aps,epsf]{revtex}
\begin{document}
\ifx\href\undefined
\else\errmessage{Don't use hypertex}
\fi

\twocolumn[\hsize\textwidth\columnwidth\hsize\csname
           @twocolumnfalse\endcsname

\title{Local Electronic Structure of Defects in 
Superconductors}
\author{Michael E. Flatt\'{e}}
\address{Department of Physics and Astronomy, 
University of Iowa, Iowa City, Iowa 52242}
\author{Jeff M. Byers} 
\address{Materials Physics, Naval Research Laboratory, Washington D.C. 
20375 } 
\date{December 16, 1996}
\maketitle
\begin{abstract}
The electronic structure near defects
(such as impurities)
in superconductors is explored using a new, fully self-consistent
technique.  This technique exploits the short-range
nature of the impurity potential and the induced change in
the superconducting order parameter to calculate features in
the electronic structure down to the atomic scale with
unprecedented spectral resolution. Magnetic and non-magnetic
static impurity potentials are considered, as well as
local alterations in the pairing interaction. Extensions
to strong-coupling superconductors and superconductors with
anisotropic order parameters are formulated.

\end{abstract} 
\vskip 2pc ] 

\narrowtext

\section{Introduction}

Low-temperature superconductors almost always have a high
concentration of non-magnetic impurities. Even in the dirty
limit, however, where the mean free path is shorter than the
coherence length, superconductivity endures\cite{PALee}.
This phenomenon can be understood by generalizing BCS pairing\cite{BCS}
to be between degenerate
Kramers partnered states in a time-reversal invariant
system\cite{AndersonKS}.  Magnetic impurities, which do break
time-reversal invariance, have more profound effects on the
superconductor in dilute concentrations than non-magnetic impurities,
lowering the critical temperature T$_c$\cite{Mathiass,Herring,Suhl}
and producing localized states within the 
gap\cite{Yu,Shiba,Rusinov,MHZ1,MHZ2,MHZ3,Kondo}
which 
at sufficient concentrations hybridize to produce gaplessness\cite{AG}. 
In the course of the investigation of the effects of
impurities on superconductivity during the last four
decades, the primary emphasis has been 
the influence of impurities on bulk properties.
These effects have been
treated within a strong-coupling formalism (e.g. Ref.~\onlinecite{Carbotte}),
but only very recently self-consistently and beyond the Born
approximation\cite{Jarrell}.
The above work was primarily concerned with bulk or impurity-averaged
characteristics and ignored
the spatial structure of electronic properties near to the impurity.

Among the first local properties calculated in the vicinity of an
impurity in a superconductor
were the structures of screening clouds around a
charged impurity\cite{Prange,Hurault} and a magnetic
impurity\cite{Hurault,Anderson}
in a superconductor (characterized by exponentially-decaying
Friedel-like\cite{Friedel} oscillations). 
The oscillation of the order
parameter around a magnetic impurity was first evaluated
without
self-consistency\cite{Tsuzuki,Heinrichs,Kummel}.
A self-consistent calculation
of the order parameter at the impurity and very far away
for {\it weak} impurity potentials was reported by
Schlottmann\cite{Schlottmann} for magnetic impurities and
Fetter\cite{Fetter} for non-magnetic impurities.

Interest in local properties near impurities in
superconductors has been
revived by the capability of scanning
tunneling microscopy (STM) to perform localized spectroscopic
measurements. 
The spatially-dependent differential conductivity around a non-magnetic 
impurity
at the surface of a superconductor
has been considered theoretically\cite{BFS} for both
isotropic and anisotropic order parameters. This
differential conductivity is proportional to the local
density of states (LDOS) around the impurity.  In Ref.~\onlinecite{BFS} the
impurity was modeled as a point defect, 
and spatial oscillations in the LDOS
at various voltages were calculated. 
These oscillations are the superconductor's analog of
oscillations in the LDOS
created by step edges and defects on
noble metal surfaces\cite{Avouris,Crommie}. 
The superconductor's LDOS oscillations would
allow one to measure the anisotropy of the superconductor's order parameter.
The conductance oscillations for
voltages just above a gap minimum or maximum are strongly
pronounced in the real-space directions corresponding to the
momenta of the gap minimum and maximum. 
Calculations followed which
considered sharp energy features in the scattering process,
such as resonant states\cite{Choi,Balatsky}.
Nevertheless, an important assumption of these calculations
has remained unchecked in detail, that the electronic distortions
induced by the 
impurity are local, including the deformation of the
order parameter. 
Self-consistent calculations using the Bogoliubov-de Gennes
(BdG) equations\cite{deGennes} followed for two-dimensional
systems\cite{Xiang,Franz,Onishi,Flatunp}, but 
have been hampered by finite-size effects.

A magnetic impurity differs from a non-magnetic impurity in
that a localized state exists around 
it\cite{Yu,Shiba,Rusinov,MHZ1,MHZ2,MHZ3,Kondo,noteus}.
The first calculations of the LDOS of the localized and
continuum states around a magnetic impurity were performed
recently both with a simplified analytic model and
numerically
via the new self-consistent technique\cite{Flatteshort} described in this paper.
Calculations of the LDOS of the localized state with angular
momentum quantum number $\ell=0$ were performed with 
a slightly different analytic model and numerically in two
dimensions via the
self-consistent BdG equations\cite{SalkBalat}. These
calculations were motivated by preliminary experimental
results around a Gd atom on a niobium
surface\cite{Yazdani}.

Another type of defect is a local change in the
pairing interaction. The resulting spatially-dependent order
parameter can then distort the LDOS.  An order parameter
suppression can even localize states, as in a vortex
core\cite{Caroli,Bardeen}.

The local electronic properties of all of these defects
can be calculated self-consistently from the
Gor'kov equation\cite{AGD} without further
approximation with the new technique introduced 
in Ref.~\onlinecite{Flatteshort}.
Our technique for calculating the
electronic structure around a defect in a superconductor
is related to the Koster-Slater inversion
techniques for determining the local electronic structure of impurities
in metals\cite{Koster1,Koster2}.
Since its
original application to localized vibrational
modes\cite{phonons}, this algorithm has been applied to
numerous problems including 
deep levels in semiconductors\cite{deepbook}
and impurity states in magnets\cite{Wolfram}.
The new Koster-Slater technique separates space around the
defect into two regions: the near field and the far field. The
far field is a region distant enough from the defect that
the potential is insignificant and the order parameter has
relaxed back to its homogeneous value. The near field is the
region close to the defect where the
potential is finite or the order parameter is distorted. 
In essence, the Gor'kov
equation that determines the Green's functions of the 
inhomogeneous superconductor
is inverted in the real space region of the near field.
This paper describes the technique in detail, expands
on an analytic model introduced in
Ref.~\onlinecite{Flatteshort} and reports several calculations of
the properties of the defects described above. 

Previous attempts to calculate the 
structure of the near field have often used
the BdG equations\cite{deGennes}.  
Other theories of inhomogeneities, such as Ginzburg-Landau
theory\cite{LG} or the Eilenberger
equations\cite{Eilenberger}, treat the spatial degrees of
freedom as coarse-grained over the superconductor's
coherence length.
Coarse-grained approximations are not appropriate
for considering electronic structure on the atomic scale near
a defect. 
The BdG equations are 
generalized Schr\"odinger equations for the electron and
hole wavefunctions of a quasiparticle, and are valid for
a superconductor with an arbitrarily-varying order parameter, 
only constrained by the validity of BCS theory. 

Unfortunately, these equations have
significant practical difficulties as well. Despite qualitative
success modeling STM measurements of a single vortex in
superconducting NbSe$_2$\cite{Hess}, 
calculations of the electronic 
structure\cite{Shore,Overhauser1,Overhauser2,Gygi1,Gygi2} 
using the BdG equations
are hampered by the difference in
energy scales between the Fermi energy and the order
parameter.  Since the BdG equations are solved numerically
for a finite system, the difficulty of the calculation is
determined by the necessary spectral resolution. The key
energy scale which must be resolved is the superconducting
gap. Thus, the numerical difficulty increases as the ratio of the 
Fermi energy to the gap becomes large.
Hence the band structure assumed for the
superconductor must be somewhat unrealistic (for
Refs.~\onlinecite{Gygi1,Gygi2} the Fermi wavelength was approximately
100\AA, which is inappropriately large\cite{NbSe2bs}).
This limitation extends to calculations of 
the interaction between a vortex and an impurity\cite{guinea},
the characteristics of the vortex lattice\cite{latticee,latticet}, and
work on a non-magnetic impurity\cite{Xiang,Franz,Onishi,Flatunp} and
a magnetic impurity\cite{SalkBalat} in a
two-dimensional $s$-wave or $d$-wave superconductor.
In contrast, the computational requirements of the 
Koster-Slater technique are determined by the range of
the impurity potential, rather than the necessary spectral precision.

In the Section II of this paper we first describe the BdG
formalism for local defect potentials and then compare
with the new, Koster-Slater formalism. 
Section III describes an analytic model, based on a
delta-function potential, which reproduces some of the
quantitative behavior of the numerical results.
Section IV discusses the results of
the numerical calculations for magnetic
impurities, non-magnetic impurities, impurities
incorporating both magnetic and non-magnetic potentials, and
inhomogeneities in the pairing interaction.
A heuristic picture of the electronic 
structure near these impurities will be presented here, and
the calculations will be compared with the analytic model of
Section III.  Section V generalizes the formalism of Section II to the
case of 
strong-coupling and anisotropically-paired superconductors.

\section{Formalism}
\subsection{Bogoliubov-de Gennes Equations}

To place our new formalism in context we will contrast it with the BdG
equations, which are  Schr\"odinger-like equations for the electron
and  hole components of the quasiparticle wavefunctions $u({\bf x})$
and $v({\bf x})$ respectively.  These are, for a free
electron band structure with mass $m$, the positive-energy ($E$) solutions to
\begin{eqnarray}
\left[-{(\hbar\nabla)^2\over 2m} -E + V_{0}({\bf x}) + \sigma
V_S({\bf x}) \right]u_\sigma({\bf x}) + \Delta({\bf x})v_\sigma({\bf
x}) &&=  0\nonumber\\ \left[{(\hbar\nabla)^2\over 2m} -E - 
V_{0}({\bf x}) + \sigma V_S({\bf x}) \right]v_\sigma({\bf 
x}) + \Delta({\bf x})u_\sigma({\bf
x}) &&= 0.\nonumber\\ \label{bdg}
\end{eqnarray}
Here $\sigma V_S$ is a position-dependent, spin-dependent 
potential, such as one originating from 
an impurity with a classical spin.
$V_0$ is a position-dependent non-magnetic potential and 
$\Delta({\bf x})$ is the inhomogeneous order parameter. 
$\Delta({\bf x})$ can be chosen real since the defect
potential is real\cite{topol}.
The quantization direction of
the electronic spins in the superconductor ($\sigma = \pm 1/2$)
is chosen parallel
to the classical spin. A classical spin has no quantum dynamics,
and cannot flip the quasiparticle spin. Hence
spin is a good quantum number for the quasiparticles and
only two coupled equations (Eqs.~(\ref{bdg})) are required.
The combinations $\sigma V_S\pm V_0$ have physical significance:
$\sigma V_S+V_0$ is the potential felt by an electron of
spin $\sigma$, while $\sigma V_S-V_0$ is the potential felt
by a hole of spin $\sigma$. 

The spatially-dependent order parameter is 
determined self-consistently:
\begin{equation}
\Delta({\bf x}) = \sum_{n\sigma} \gamma({\bf x},E_{n\sigma})
u_{n\sigma}({\bf x})v^*_{n\sigma}
({\bf x}){\rm tanh}\left({ E_{n\sigma}\over 2k_BT}
\right)\label{bdgsc} \end{equation}
where $n$ labels the states for each spin $\sigma$, $T$ is the temperature,
 and $k_B$ is Boltzmann's
constant. 
$\gamma({\bf
x},E_{n\sigma})$
is the effective electron-electron interaction potential, which
is
\begin{eqnarray}
\gamma({\bf x},E_{n\sigma}) &&=
\gamma_o\qquad E_{n\sigma}<\hbar\omega_D\nonumber\\
&&= 0\qquad E_{n\sigma}>\hbar\omega_D.
\end{eqnarray}

For a spherically-symmetric 
defect the wavefunctions are 
eigenstates of the angular momentum with quantum 
numbers $\ell$ and $m$.  
Typically the defect is placed in a sphere
of radius $R$ with appropriate boundary conditions. 
The value of $R$ is determined by 
the spectral resolution necessary for accurately 
evaluating Eq.~(\ref{bdgsc}) and the spectral width of features
measurable by (for example) the STM.  The
typical complications resulting from approximating an infinite 
system by a finite-size system apply, such as 
discrete states above the energy gap and the 
heavy investment of computer time required for 
large values of $R$. For example, in the calculations for
the vortex in NbSe$_2$\cite{Gygi1,Gygi2}, $\epsilon_F/\Delta_o = 32$ 
was the largest
ratio of the Fermi energy to the homogeneous order parameter 
considered. This value
is unrealistic, and is a result of inappropriately fitting the coherence
length and upper critical field of NbSe$_2$ to a
free-electron model. A more realistic band
structure\cite{NbSe2bs} has a bandwidth to energy gap ratio
at least an order of magnitude greater, and a highly
anisotropic Fermi velocity.
A system where the free-electron model is appropriate,
such as niobium, has an $\epsilon_F/\Delta_o = 705$ and is numerically
inaccessible for general potentials $V_S$ and $V_0$.

\subsection{Koster-Slater Inversion Formalism}

We now introduce a self-consistent method which works within a 
sphere whose radius is determined by the range of the 
defect's potential and utilizes the 
continuum spectrum of the homogeneous 
superconductor. In essence, we invert the Gor'kov
equation in real space. The Gor'kov equation\cite{AGD}
for a defect in a superconductor
can be written in the Nambu formalism\cite{Schrieffer} as: 
\begin{eqnarray}
\int d{\bf x''} \left[ {\bf \delta}({\bf x}-{\bf x''}) - 
{\bf g}({\bf x},{\bf x''}; \omega){\bf V}({\bf 
x''})\right] &&{\bf G}({\bf x''},{\bf x};\omega) \nonumber\\
&&= {\bf g}({\bf x},{\bf x'};\omega)\label{KS}
\end{eqnarray}
where the inhomogeneous retarded Green's function,
\begin{equation}
{\bf G}({\bf x},{\bf x'};\omega) = \left(\begin{array}{cc}G_\uparrow({\bf
x},{\bf x'};\omega)&F({\bf x},{\bf x'};\omega)\\
F^*({\bf x},{\bf x'};\omega)& -G_{\downarrow}^*({\bf x},{\bf x'};-
\omega)\end{array}\right). \label{nambuG}\end{equation}

The elements of this matrix are
\begin{eqnarray}
G_\uparrow({\bf x},&&{\bf x'};\omega) =\nonumber\\
&& -i\int_{-\infty}^\infty dt
{\rm e}^{i\omega t}\theta(t)
\langle 0| \{ \psi_\uparrow({\bf x'};t),\psi_\uparrow^\dagger
({\bf x};0)\}|0\rangle ,\\
F({\bf x},&&{\bf x'};\omega) =\nonumber\\&& -i\int_{-\infty}^\infty 
dt {\rm e}^{i\omega t}\theta(t)
\langle 0| \{ \psi_\uparrow({\bf x'};t),\psi_{\downarrow}
({\bf x};0)\}|0\rangle ,\\
F^*({\bf x},&&{\bf x'};\omega) =\nonumber\\&& -i\int_{-\infty}^\infty 
dt{\rm e}^{i\omega t}\theta(t)
\langle 0| \{ \psi_{\downarrow}^\dagger({\bf x'};t),
\psi_{\uparrow}^\dagger({\bf x};0)\}|0\rangle ,\\
-G^*_{\downarrow}({\bf x},&&{\bf x'};-\omega) =\nonumber\\&& 
-i\int_{-\infty}^\infty dt{\rm e}^{i\omega t}\theta(t)
\langle 0| \{ \psi_{\downarrow}^\dagger({\bf x'};t),\psi_{\downarrow}
({\bf x};0)\}|0\rangle .
\end{eqnarray}
The explicit subscripts $\uparrow$ and $\downarrow$ do {\it not}
refer to the spin of the excitation in the superconductor but
rather to the spin band of the normal state used to
construct the excitation. The key concept is that the
spin-{\it up} band contains both {\it up} electrons
and {\it down} holes just as the spin-{\it down} band contains both
{\it down} electrons and {\it up} holes. 
The convention employed here
is standard in the theory of semiconductors where a spin-up electron
excited above the Fermi energy leaves a spin-down hole below
the Fermi energy.  This is convenient for
magnetic potentials since if spin-up electrons are attracted
to a magnetic impurity spin-down holes should be repelled by
the impurity. 
In the presence of a single classical impurity spin, the
quasiparticle spin is a good quantum number despite
electron-hole mixing. Our convention determines the
composition of a spin-up quasiparticle  to be part spin-up
electron and part spin-up hole, rather than part spin-up
electron and part spin-down hole.

For $\omega>0$, ${\bf G}({\bf x},{\bf x'};\omega)$ describes spin-up 
excitations, involving the mixing
of electrons in the spin-up band with holes in the spin-down band. 
For $\omega<0$,
${\bf G}({\bf x},{\bf x'};\omega)$ describes spin-down excitations, 
involving the mixing
of electrons in the spin-down band with holes in the spin-up band.
Since spin is a good quantum number it is not necessary to
use a $4\times 4$ formalism, such as that of
Ref.~\onlinecite{Ambegaokar}. The notation here has been
simplified and improved relative to Ref.~\onlinecite{Flatteshort}.

The homogeneous Green's function ${\bf g}$ is independent of $\sigma$, so
\begin{equation}
{\bf g}({\bf x},{\bf x'};\omega) = \left(\begin{array}{cc}g({\bf
x},{\bf x'};\omega)&f({\bf x},{\bf x'};\omega)\\
f^*({\bf x},{\bf x'};\omega)& -g^*({\bf x},{\bf x'};-
\omega)\end{array}\right). \end{equation}
The potential
\begin{equation}
{\bf V}({\bf x''}) = \left( 
\begin{array}{cc}
V_S({\bf x''})+ 
V_{0}({\bf 
x''})&\delta\Delta({\bf x''})\\ 
\delta\Delta({\bf x''})& V_S({\bf 
x''})- V_{0}({\bf x''})\end{array}
\right)\label{potn}
\end{equation}
where $\delta\Delta({\bf x}) = 
\Delta({\bf x})-\Delta_o$ and 
$V_S$, $V_{0}$, and  $\Delta({\bf x})$ 
have similar meaning above as in the BdG
equations. The factor of one-half from the electron spin has been
incorporated into the potential $V_S$. To be concrete, and
without loss of generality, the spin direction attracted by the 
potential will be called spin up, and the spin direction
repelled will be called spin down\cite{FMAFM}.
The self-consistency equation for the order parameter is

\compage

\twomini{
\begin{equation}
\delta\Delta({\bf x}) = \int_{-\infty}^\infty d\omega \gamma
({\bf x},\omega)n(\omega)
{\rm Im}F({\bf x},{\bf x};\omega) - \Delta_o\label{KSsc}
\end{equation}
where $n(\omega)$ is the Fermi occupation function at
temperature $T$.

Self-consistent spin-dependent and charge-dependent potentials can
also be constructed for ${\bf V}({\bf x})$ using the calculated spatial
structure of the spin $s({\bf x})$ and charge $\rho({\bf x})$ around the 
defect, 

\lftlne} {
\begin{eqnarray}
s({\bf x}) =&& \int_{-\infty}^\infty d\omega n(\omega) 
\left(-{ {\rm Im}
\left[G_\uparrow({\bf x},{\bf x};\omega) -G_\downarrow({\bf x},
{\bf x};\omega)\right]\over \pi }\right) \nonumber\\&&\\
\rho({\bf x}) =&& \sum_\sigma\int_{-\infty}^\infty d\omega n(\omega) 
\left(-{{\rm Im}G_\sigma({\bf x},{\bf x};\omega) \over \pi}\right).
\end{eqnarray}
No calculations are
reported for such potentials\cite{deGennes} in this paper.
}

For a BCS
superconductor with an isotropic gap in a parabolic band,
for $\omega$ much smaller than the Fermi energy,

\begin{eqnarray}
g({\bf x},{\bf x'};\omega) =&&-{\pi N_o\over k_Fr} {\rm
e}^{-\sqrt{\Delta_o^2-\omega^2}r/\pi\Delta_o\xi}\left(\cos k_Fr + {\omega\over
\sqrt{\Delta_o^2-\omega^2}}\sin k_Fr\right)\nonumber\\
f({\bf x},{\bf x'};\omega) =&&-{\pi \Delta_o N_o\over
k_Fr\sqrt{\Delta_o^2-\omega^2}}{\rm e}
^{-\sqrt{\Delta_o^2-\omega^2}r/\pi\Delta_o\xi}\sin k_Fr
\label{bareg}\end{eqnarray} 

\twomini{
\vskip\baselineskip

\noindent where $r = |{\bf x}-{\bf x'}|$ and $N_o$ is the density of states for
each spin at
the Fermi level.
The coherence length, $\xi = \hbar v_F/\pi\Delta_o$,  where $v_F$ is
the Fermi velocity. 
These expressions are valid for $\omega$ above and
below $\Delta_o$ so long as the imaginary parts of both $f$ and $g$
are multiplied by ${\rm sgn} \omega$. 

One strength of this formalism is its reliance on the 
short-range nature of the defect's potential. 
Solution of Eq.~(\ref{KS}) requires inverting the frequency-dependent 
real-space matrix 
\begin{equation}
{\bf M}({\bf x},{\bf x'};\omega) = {\bf \delta}({\bf x}-{\bf x'}) -
{\bf g}({\bf x},{\bf  x'};\omega) {\bf V}({\bf x'}).
\end{equation}
The structure of ${\bf M}({\bf x},{\bf x'};\omega)$ allows
for a precise description of the difference between the
near field and the far field.  We require that the defect's
potential ${\bf V}({\bf x})$ is zero for $|{\bf x}|>R$. The space
$|{\bf x}|>R$ belongs to the far field, whereas the space
$|{\bf x}|\le R$ belongs to the near field.
We can then separate any real-space matrix ${\bf A}$ symbolically
into four pieces:
\vskip.75pc

\lftlne }{ \rgtlne
\begin{equation}
{\bf A} = \left(\begin{array}{cc}
{\bf A}^{n\rightarrow n}& {\bf A}^{n\rightarrow f}\\
{\bf A}^{f\rightarrow n}& {\bf A}^{f\rightarrow f}\end{array}\right)
\end{equation}
where $n$ and $f$ label the near-field region and far-field
region, respectively. 
The particular example of ${\bf M}$ is block-triangular:
\begin{eqnarray}
{\bf M} =&& \left(\begin{array}{cc}
{\bf I} - {\bf g}^{n\rightarrow n}{\bf V}&\qquad0\\
-{\bf g}^{f\rightarrow n}{\bf V}&\qquad{\bf I}\end{array}\right)\nonumber\\
\nonumber\\
{\bf
M}^{-1} =&& \left(\begin{array}{cc}
({\bf I}-{\bf g}^{n\rightarrow n}{\bf V})^{-1}&\qquad0\\
{\bf g}^{f\rightarrow n}{\bf V}({\bf I}-{\bf g}^{n\rightarrow
n}{\bf
V})^{-1}&\qquad{\bf I}\end{array}\right).\label{Mschem}
\end{eqnarray}
It is clear from Eq.~(\ref{Mschem}) that the computational
effort in inverting ${\bf M}$, and thus finding the
inhomogeneous electronic structure, is entirely determined by
the complexity of inverting ${\bf M^{n\rightarrow n}}$.
We also obtain the useful
result that the electronic structure in the far field is
easily determined once the electronic structure in the near
field is known.

The local density of states in the near field is merely
}
\begin{eqnarray}
-{1\over \pi}\sum_\sigma{\rm Im} G_{\sigma}({\bf x},{\bf x};\omega) = 
-{1\over \pi}{\rm Im} \Big\{2&&g({\bf x},{\bf x};\omega) + 
\int d{\bf x'}\left[\left[ ({\bf M}^{n\rightarrow n})^{-1}({\bf x},{\bf
x'};\omega)\right]{\bf g}({\bf x'},{\bf x};\omega)\right]_{11}\nonumber\\
- &&
\int d{\bf x'}\left[\left[ ({\bf M}^{n\rightarrow n})^{(-1)*}({\bf x},{\bf
x'};-\omega)\right]{\bf g}^*({\bf x'},{\bf x};-\omega)\right]_{22}
\Big\}
\end{eqnarray}
while in the far field it is

\begin{eqnarray}
-{1\over \pi}\sum_\sigma{\rm Im}&& G_{\sigma}({\bf x},{\bf x};\omega) =
-{1\over \pi}{\rm Im} \Big\{2g({\bf x},{\bf x};\omega) 
\nonumber\\
&&+\int d{\bf x'} d{\bf x''}
\left[{\bf g}({\bf x},{\bf x'};\omega){\bf V}({\bf x'})
\left[({\bf M}^{n\rightarrow n})^{-1}({\bf x'},{\bf
x''};\omega)\right]{\bf g}({\bf x''},{\bf x};\omega)\right]_{11}
\nonumber\\
&&-\int d{\bf x'} d{\bf x''}
\left[{\bf g}^*({\bf x},{\bf x'};-\omega){\bf V}({\bf x'})
\left[({\bf M}^{n\rightarrow n})^{(-1)*}({\bf x'},{\bf
x''};-\omega)\right]{\bf g}^*({\bf x''},{\bf x};-\omega)\right]_{22}
\Big\}.
\label{grr}
\end{eqnarray}
\break

\noindent
In the limit that the size of the near-field approaches a
point, Eq.~(\ref{grr}) is 
\begin{equation}
-{1\over \pi}\sum_\sigma{\rm Im} G_{\sigma}({\bf x},{\bf x};\omega) =
-{1\over \pi}{\rm Im} \Big\{2g({\bf x},{\bf x};\omega) 
+ \left[{\bf g}({\bf x},{\bf 0};\omega){\bf V_{eff}}(\omega)
{\bf g}({\bf 0},{\bf x};\omega)\right]_{11}
- \left[{\bf g}^*({\bf x},{\bf 0};-\omega){\bf V^*_{eff}}(-\omega)
{\bf g}^*({\bf 0},{\bf x};-\omega)\right]_{22}
\Big\} .  \label{grreff}
\end{equation}
The inhomogeneous part of the right-hand side
of Eq.~(\ref{grreff}) has the following form if the
off-diagonal elements of ${\bf V_{eff}}(\omega)$ can 
be ignored (this is appropriate when self-consistency is
not important):
\begin{equation}
-{1\over \pi}{\rm Im}\left\{\left(
[{\bf V_{eff}}(\omega)]_{11} - 
[{\bf V_{eff}}^*(-\omega)]_{22} \right)\left(
g({\bf x},{\bf 0},\omega)
g({\bf 0},{\bf x},\omega) -  f({\bf x},{\bf
0},\omega)f({\bf 0},{\bf x},\omega)\right)\right\}.\label{nonB}
\end{equation}
\twomini{
\vskip.3pc

\noindent
Hence the spatially-dependent features reported in Ref.~\onlinecite{BFS} are
qualitatively retained in our more complete model.
For a magnetic potential in the Born approximation
there would be no such signal,
since
\begin{equation}
{\bf V_{eff}}(\omega) = \left(
\begin{array}{cc}V_S&0\\
0&V_S \end{array}\right)
\end{equation}
so the distortions in the LDOS of the two spin directions
are equal and opposite.

The local density of states under ideal conditions
is directly measured by a
scanning tunneling microscope. The differential conductivity 
measured at a point ${\bf x}$, voltage $V$ and
temperature $T$ can be related
to the local density of states at the tip location as follows:
\begin{equation}
{dI({\bf x},V,T)\over dV} = {e^2\over h} \int_{-\infty}^\infty
d\omega {\partial n_{STM}(\omega)\over \partial\omega}\sum_\sigma
\left({{\rm Im}G_\sigma({\bf x},{\bf x};\omega)\over \pi}\right).
\label{STMLDOS}
\end{equation}
Here $e$ is the charge of the electron and $n_{STM}(\omega)$ is the 
Fermi function
in the STM tip 
\begin{equation}
n_{STM}(\omega) = \left[1+\exp{\left({\omega - 
eV\over k_BT}\right)}\right]^{-1}.
\end{equation}
The local density of states is proportional to the imaginary part of the
retarded Green's function fully dressed by the interaction of the
electronic system with the impurity.

Certain features of the above equations simplify their
numerical implementation.
Each angular momentum channel  
constitutes an independent block-diagonal
submatrix in ${\bf M}(\omega)$. Since the bare Green's functions in
Eqs. (\ref{bareg}) have analytic expansions in spherical harmonics,
${\bf M}(\omega)$ can be calculated quickly. These
expansions are detailed in the Appendix.

The value of $R$ is governed by the longest-range 
potential. In this paper that will be 
$\delta\Delta({\bf x})$. We model ${V_S}$ and 
$V_{0}$ with Gaussians of radius $k_F^{-1}$ and evaluate
Eq.~(\ref{KSsc}) at $T=0$.  The 
$\delta\Delta({\bf x})$ for 
$v_s=\pi N_o| \int d{\bf x}V_S({\bf x})| = 1.75$ and
$v_0=\pi N_o| \int d{\bf x}V_0({\bf x})| = 0$
 is shown in Figure~1.
While 
oscillating
with wavelength $\sim \pi k_F^{-1}$, $\delta\Delta({\bf x})$
falls off to a 
negligible potential within $10k_F^{-
1}$. A typical radial grid 
\vskip\baselineskip
}{ \rgtlne

\noindent
of 100 points provides a 
numerically robust solution. The self-consistent solution
depends on the value of 
\begin{equation}
N_o\Delta_o/k_F^3 =
(2\pi^3\xi k_F)^{-1} = (4\pi^2 \epsilon_F/\Delta_o)^{-1},
\end{equation}
 which for niobium is
$3.6\times 10^{-5}$ (Ref. \onlinecite{NbDOS}). 
This is the single
dimensionless parameter required to parametrize the 
free-electron model of a superconductor.

In contrast to the Koster-Slater technique described here,
which exploits a physical distinction between the near field
and the far field to accelerate the numerical calculation,
the BdG equations treat the
near-field and the far-field on the same level.
A large $R$ is desired for decent spectral resolution, but
the possible size of $R$ is limited by numerical
constraints. Hence a numerical implementation of the 
BdG equations, by
comparison, is typically slower and substantially less
accurate than the Koster-Slater technique.

\begin{figure}
\epsfxsize=2.25in
\epsfysize=1.30in
\epsffile{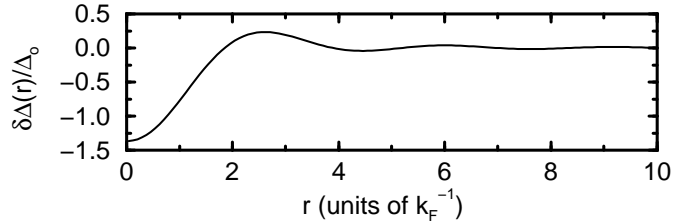}
\caption{Change in the local order parameter, $\delta\Delta({\bf x})$,
for a magnetic potential strength $v_s = N_o\int d{\bf
r}V_S({\bf r}) = 1.75$ as a function of the distance from the 
impurity $r$. The change becomes negligible beyond $10k_F^{-1}$. }
\end{figure}

\section{Analytic Solution of the Point Potential}

\subsection{Magnetic and Non-Magnetic Point Potentials}

Approximating the local potential by a delta
function
\begin{equation}
{\bf V}({\bf x}) = {\bf V}\delta({\bf x}) =
\left(\begin{array}{cc} V_S + V_0&0\\
0&V_S-V_0\end{array}\right)\delta({\bf x})
\end{equation}
leads to a simple expression for Eq.~(\ref{KS}),
}
\begin{equation}
\left(\begin{array}{cc}
1-g({\bf 0},{\bf 0};\omega)(V_S+V_0)& -f({\bf 0},{\bf 0};\omega)(V_S-V_0)\\ 
-f^*({\bf 0},{\bf 0};\omega)(V_S+V_0)& 1+g^*({\bf 0},{\bf 0};-\omega)(V_S-V_0)
\end{array}\right) 
{\bf G}({\bf 0},{\bf 0};\omega) =
 {\bf M}^{n\rightarrow n}(\omega) {\bf G}({\bf 0},{\bf 0};\omega)
= {\bf g}({\bf 0},{\bf 0};\omega). \label{analsol}
\end{equation} 
\twomini{
\vskip\baselineskip
\noindent
In principle ${\bf M}(\omega)$ can be found from the Green's
functions in Eq.~(\ref{bareg}), however there is a divergence
in the real part of $g(r;\omega)$ as $r\rightarrow 0$. This
divergence is coped with in Ref.~\onlinecite{Shiba} by discarding
the divergent piece. This approximation is essentially an
assumption of {\it strict} particle-hole symmetry (not merely linearizing
$\epsilon(k)$ around $\epsilon_F$). We now discuss the
effects of this approximation on the local properties of the
system in the {\it normal} state. The heuristics will be
simpler 
for the normal state, but the conclusions
also apply to the superconducting
state.

\subsection{Particle-Hole Symmetry in the Normal State}

In order to focus on the spin-dependent potential, 
the non-magnetic potential will be set to zero. 
The normal state properties
can be obtained from Eq.~(\ref{bareg}) for $\Delta_o=0$. That
yields a Green's function appropriate for an outgoing wave:
\begin{equation}
g({\bf x},{\bf x'};\omega) = -{\pi N_o\over k_Fr} {\rm
e}^{ik_Fr},\label{gnorm}
\end{equation}
where $\omega$ is considered close to the Fermi surface so
the change in momentum due to $\omega\ne 0$ is negligible.
The {\it inhomogeneous} local density of states for a
delta-function potential using this Green's
function is unphysical, since it diverges at $r=0$. Now
we examine the local density of states when the divergent
real part of $g$ is ignored.
\begin{equation}
-{1\over \pi}{\rm Im}G_{\sigma}({\bf x},{\bf x};\omega) = 
N_o\left(1+\left[{(\pi N_o V_S)^2\over 1-(\pi N_o
V_S)^2}\right]{\sin^2 k_Fr\over (k_Fr)^2}\right).\label{LDOSph}
\end{equation}
This expression yields the pathological behavior that the
local density of states near a spin-dependent potential
is exactly the same for spin-up
electrons as for spin-down 
\noindent electrons. 
A spatial response in the LDOS to a spin-dependent potential
that is identical for up and down spin electrons {\it only}
occurs for certain band structures with $\epsilon_F$ at special energies
({\it  e.g.} half-filled tight-binding models).
The local density of
states for $v_s = \pi N_o |V_S| = 0.1$ is plotted in Figure~2
as a function of distance from the potential. Although it is
somewhat distorted from its 
homogeneous value, it does not
show the spin-
dependent asymmetry of a more realistic
potential.  An exact calculation for a Gaussian of range $k_F^{-1}$
is also shown in Figure~2 for comparison.

A more realistic approach to coping
with the divergence in Eq.~(\ref{gnorm}) without yielding
the pathological result of Eq.~(\ref{LDOSph}) is to average
the 
real part of $g$ over a range given by the assumed range
of the potential. That yields a finite value for the local
density of states at the potential, but does not control the
behavior for small $r$. 
}{\vskip0.3pc
\rgtlne

\noindent
To perform that task, we consider a
``muffin-tin'' Green's function. This function has the form
\begin{eqnarray}
g({\bf x},{\bf x'};\omega) &&= -i{\pi N_o\over k_Fr} {\rm
e}^{i k_Fr}
\qquad\qquad r > R_o\nonumber\\
&&= -i{\pi N_o} -\pi \alpha\qquad\qquad  r<R_o \label{gKS}
\end{eqnarray}
where $\alpha$ is the average of the 
real part of $g$ over the range of the
potential, and $R_o$ is chosen so that the spatially-integrated
spectral weight of the Green's function is unchanged (as required
by probability conservation).
We show in Figure~2 the local density of states calculated
with this Green's function. 
In particular, the asymmetry in the response of spin-up
electrons and spin-down electrons to a spin-dependent
delta-function potential is governed by this
phenomenological parameter $\alpha$.
The agreement with the exact
solution is good at the origin and far from the impurity.
The muffin-tin Green's function is discontinuous,
unfortunately, but yields a better approximation of the
response of the system than the 
particle-hole symmetric
approximation (discarding the real part of the 
Green's function).

It is also possible to generate an asymmetry between\ \ 
spin-up\ \ 
and\ \ spin-down\ \ electrons\ \ by\ \ adding\ \ a\ \ non-

\vskip\baselineskip

\begin{figure}
\epsfxsize=2.5in
\epsfysize=2.00in
\epsffile{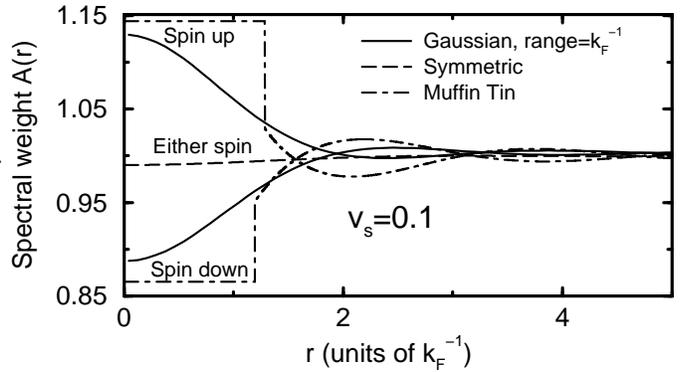}
\caption{
Spectral weight in the normal state around a magnetic impurity
with $v_s = 0.1$ as a function of the distance from the impurity $r$. 
The solid lines are exact solutions for spin-up (attracted by the
impurity)
and spin-down (repelled by the impurity)
electrons for a Gaussian
potential of range $k_F^{-1}$ in a metal with a free-electron dispersion 
relation.
The dashed line, which is the same for both spin-up and spin-down 
electrons, is
the calculated spectral weight for the particle-hole symmetric 
delta-function potential
model. The dot-dashed line is the result for a delta-function 
potential calculated
using the muffin-tin Green's functions. The muffin-tin Green's functions 
fix the
pathological result of the symmetric model that the spectral weight is 
the same for
spin-up and spin-down electrons. The muffin-tin parameter, $\alpha=0.704$, 
is determined
by the range of the Gaussian potential and is therefore not a free 
parameter. 
}
\end{figure}
}

\twomini{
\noindent 
magnetic potential
$v_0$ {\it with} the $v_s$ to parametrize the impurity, but
still maintaining a particle-hole symmetric 
band structure.  Since
the response of a particle-hole symmetric system
to a potential does not depend on
the sign 
of that potential, the $v_0$ is required to
distinguish between electrons and holes. 
An electron feels a
potential $v_s + v_0$, whereas a hole feels a potential
$v_s-v_0$.
This additional non-magnetic potential only breaks particle-hole symmetry
locally (within the 
range of the potential), whereas
for a realistic band 
structure the particle-hole symmetry 
is broken everywhere in the solid. 

For a Gaussian potential with range $a$, 
\begin{equation}
\alpha = {2\over \sqrt{\pi} k_Fa}\left( 1 + \sum_{n=2}^\infty
{1\over (2n-3)!!}\left(-{(k_Fa)^2\over 2}\right)^{n-1}\right).
\end{equation}
For the Gaussian potentials numerically calculated in this
paper, $a=k_F^{-1}$, so $\alpha = 0.704$.
\vskip-.2pc

\subsection{Self-consistency within the Analytic Model}

As seen from Figure~1, the distortion of the order parameter
is short-ranged around an impurity. We may
then consider the effect of the order-parameter distortion on
the electronic structure to be
parametrized by an effective delta-function potential
$\delta\Delta \delta({\bf x})$ similarly motivated to the
delta-function potentials for the magnetic and non-magnetic
potentials. The potential ${\bf V}({\bf x})$ is changed in the
following way:
\vskip-.1pc
\begin{equation}
{\bf V}({\bf x}) = {\bf V}\delta({\bf x}) =
\left(\begin{array}{cc} V_S + V_0&\delta\Delta\\
\delta\Delta& V_S-V_0\end{array}\right)\delta({\bf x}).
\end{equation}
The relative effect of the $\delta\Delta$ compared to the
other two potentials is likely to be small for the
potentials considered in this paper. Typically $N_oV_S/k_F^3\sim 1$
or
$N_oV_0/k_F^3\sim 1$,
and for niobium $N_o\Delta_o/k_F^{3} = 
3.6\times 10^{-5}$. Even for a small coherence length of
$\xi = 10k_F^{-1}$, $N_o\Delta_o/k_F^{3} = 1.6\times
10^{-3}$. For convenience we define $\delta_0 = \pi
N_o\delta\Delta$.
\vskip-.2pc

\subsection{Energies of localized states in the
Superconductor}

The energies of the localized states of angular momentum 
$\ell$ correspond to the positive energies $\omega_\ell =
|\Omega|$, where
\begin{equation}
{\rm det} {\bf M}^{n\rightarrow n}(\Omega) = 0,\label{detm}
\end{equation}
and the solution is traced to the $\ell$-channel block of
${\bf M}$ (see Appendix).
For the analytic model, ${\bf M}^{n\rightarrow n}(\omega)$
is the matrix shown in Eq.~(\ref{analsol}), where $g({\bf
0},{\bf 0};\omega)$ in the superconducting state is
constructed similarly to that of the normal state,
}{
\vskip-.3pc
\begin{equation}
g({\bf 0},{\bf 0};\omega) = -\pi N_o \left(\alpha + {\omega\over
\sqrt{\Delta_o^2-\omega^2}}\right).
\end{equation}
The anomalous Green's function is given by Eq.~(\ref{bareg}), since it
does not have a divergence problem as $r\rightarrow 0$,
\begin{equation}
f({\bf 0},{\bf 0};\omega) = -\pi N_o \left({\Delta_o\over
\sqrt{\Delta_o^2-\omega^2}}\right).
\end{equation}
This analytic model only has localized states in the
$\ell=0$ angular momentum channel, as expected for a delta-function
potential. Those energies are
\begin{equation}
\omega_o = \left|{v_s\delta_0 \pm \left[(2v_s\delta_0)^2
- 4(v_s^2+\gamma^2)(\delta_0-\gamma^2)\right]^{1/2}\over
v_s^2+\gamma^2}\right|\Delta_o\label{allofit}
\end{equation}
where
\begin{equation}
\gamma = \left[(1+\alpha^2)(v_s^2-v_0^2-\delta_0^2) - 2\alpha
v_0 -
1\right]/2.
\end{equation}
Eq.~(\ref{allofit}) reduces to a result obtained by Shiba\cite{Shiba}
when $v_0=\alpha=\delta_0=0$,
 a result obtained by Rusinov\cite{Rusinov}
when
$\alpha=\delta_0=0$, a result obtained by Salkola, Balatsky
and Schrieffer\cite{SalkBalat} when $\alpha=0$, and a result
obtained by us\cite{Flatteshort} when $v_0=\delta_0=0$.

One solution of Eq.~(\ref{allofit}) is a spin-up quasiparticle and 
the other is a spin-down
quasiparticle. There may be only one real solution to 
Eq.~(\ref{allofit}); then
only one $\ell =0$ localized state exists
around the impurity. This occurs for large $v_s$.
When $v_s=0$ the localized states are due to 
order parameter suppression, and
the energies of the two spin states are
degenerate. This follows from time-reversal symmetry  in the
absence of a magnetic potential.
For small $v_s$ the two energies are split by
an amount
\begin{equation}
\Delta\omega_o = {2v_s\delta_o\Delta_o\over v_s^2+\gamma^2}.
\end{equation}

\subsection{Spectral Weight Asymmetry in the Analytic Model}

A spin-up quasiparticle consists of amplitudes for
a spin-up electron (electron
in a spin-up state), and a spin-up hole (electron missing
from a spin-down state).  Therefore the spectral weight of 
a spin-up localized state will be divided between an electron-like
pole in the spin-up band at $\omega=\omega_o$ 
(with weight $A_\uparrow({\bf r};\omega_o)$)
and a hole-like pole in the spin-down band at
$\omega=-\omega_o$
(with weight $A_\downarrow({\bf r};-\omega_o)$).
These two types of excitation are
independently resolvable by a scanning tunneling microscope
since at positive sample voltage relative to the tip,
the STM places electrons in the sample, whereas at negative
}

\widetext

\noindent
sample voltage the STM places holes in the sample. We define the energy of the 
pole in the spin-up band to be $\omega_\uparrow$ and in the spin-down
band to be $\omega_\downarrow$. Even though $\omega_o$ is always positive,
$\omega_\uparrow$ can be positive or negative, and 
$\omega_\uparrow=-\omega_\downarrow$.

The spatial structure of the 
spectral weights of the spin-up band and
spin-down band components of the localized state 
are given by 
\vskip1pc
\begin{eqnarray}
A_\sigma({\bf r};\omega) =&& {\pi N_o\Delta_o\over 2v_s}\delta(\omega-
\omega_\sigma)
{\sqrt{\Delta_o^2-\omega^2}\over\Delta_o}\bigg[ 
{(v_0-v_s)\Delta_o^2+(v_0+v_s)\omega^2+(v_0+v_s)\alpha^2(\Delta_o^2-\omega^2)
\over \Delta_o^2}\nonumber\\
&&
+{2(v_0+v_s)\alpha\
\omega\sqrt{\Delta_o^2-\omega^2}
-(1+\alpha^2)(v_0^2-v_s^2)\left(\alpha(\Delta_o^2-\omega^2)+
\omega\sqrt{\Delta_o^2-\omega^2}\right)
\over \Delta_o^2}\bigg]\nonumber\\
&&
\qquad\qquad
\qquad\qquad\qquad\qquad 
\qquad\qquad\qquad\qquad 
\qquad\qquad\qquad\qquad r<R_o\nonumber\\
=&& {\pi N_o\Delta_o\over (k_Fr)^2}{(\Delta_o^2-\omega^2)^{3\over 2}\over
2v_s\Delta_o^3}{\rm e}^{-{2r\over \pi\xi}\left({\sqrt{\Delta_o^2-
\omega^2}\over \Delta_o}\right)}
\delta(\omega - \omega_\sigma)
\times\nonumber\\
&& \qquad \bigg[\left(v_s {\Delta_o^2+\omega^2\over \Delta_o^2-\omega^2} 
- v_0 + (v_s^2-v_o^2)
\left\{\alpha - {\omega\over\sqrt{\Delta_o^2-\omega^2}}\right\}
\right)\sin^2 (k_Fr)\nonumber\\
&&\qquad + (v_s+v_0)\left(1+(v_s-v_0)\left\{{\omega\over
\sqrt{\Delta_o^2-\omega^2}}-\alpha\right\}\right)\cos^2(k_F r)\nonumber\\
&&\qquad + 2(v_s+v_0)\left(v_0-v_s+[1-\alpha (v_0-v_s)]{\omega\over 
\sqrt{\Delta_o^2-\omega^2}}\right)
\sin (k_Fr)\cos(k_Fr)
\bigg]
\nonumber\\
&&
\qquad\qquad
\qquad\qquad\qquad\qquad 
\qquad\qquad\qquad\qquad 
\qquad\qquad\qquad\qquad 
r>R_o\label{MFSC}
\end{eqnarray}
\vskip1.5pc
\twomini{
\vskip5.40pc
\noindent
The above expression is set up for use as a muffin-tin Green's 
function. The construction
of such a Green's function requires that the spectral weight integrates 
to one.
However, if $R_o\rightarrow 0$, the above expression can be 
integrated over all space, and still
yields one-half for the
spin-up band and one-half for the spin-down band to order
$\Delta_o/\epsilon_F$. Hence the
quasiparticle is one-half electron and one-half
hole. This provides additional confidence in the above expressions.

The frequency-integrated
weight at the defect of the two types of
excitation can be calculated within the analytic model, and is
(for $\delta_0=0$)
\begin{equation}
{A_{\uparrow}({\bf r}={\bf 0})\over A_{\downarrow}({\bf
r}={\bf 0})}
= {1+2\alpha(v_0-v_s) + (1+\alpha^2)(v_0-v_s)^2\over 
1+2\alpha(v_0+v_s)+(1+\alpha^2)(v_0+v_s)^2}
\end{equation}
For a spin-down
quasiparticle there is an electron-like pole in the spin-down 
band and a hole-like pole in the spin-up band and the
relative weight is still given by the expression above.
This expression for $v_0 = 0$ was reported in
Ref.~\onlinecite{Flatteshort}.

In the following section these results are compared with the
numerical calculations of properties in the superconducting
state {\it and} in the normal state.

}
{\rgtlne
\section{Comparison with Numerical Results and Discussion}

\subsection{Non-Magnetic Impurity}

Even strong non-magnetic impurities at moderate
concentrations
will not suppress the critical temperature of 
a superconductor\cite{AndersonKS}. Nevertheless, it was
recognized early on\cite{Fetter} that the local order
parameter may be affected. In Ref.\onlinecite{Fetter} the
effect of the non-magnetic impurity was 
calculated 
in the far field by modeling the impurity potential 
with a phase shift. The phase shifts were evaluated for two 
models: a spherical square-well potential and a delta-shell
potential. Self-consistency was ignored by only focusing on 
regions far from the
impurity where the change in the order parameter is small
compared to its homogeneous value.  The order parameter change
due to the impurity was
found to oscillate with the Fermi wavelength, and decay as
$r^{-2}$.

Figure~3 shows the spectral weight $A({\bf r};\omega)$ 
at several frequencies above the energy gap near a
strong non-magnetic impurity with a Gaussian potential 
of range $k_F^{-1}$ and strength $v_0 = 7/8$, calculated
self-consistently for
a superconductor with $\xi k_F = 449$ (niobium). The 
spectral weights are suppressed to
approximately $30\%$ of their homogeneous value at the center of the
potential. 
 Only
continuum states are shown since no localized states 

} \twomini{
\begin{figure}
\epsfxsize=2.5in
\epsfysize=2.000in
\epsffile{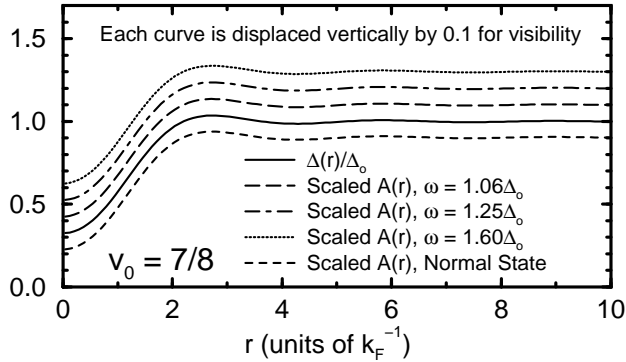}
\caption{Comparison of the spectral weight in the superconducting 
state at several
frequencies to the spectral weight of the normal state and the 
spatial dependence
of the order parameter. All of these curves are indistinguishable, 
but they have
been displaced for better visibility. The spectral weights are 
normalized to their
value at large distances from the impurity $r$, and the order 
parameter is normalized to
the homogeneous order parameter $\Delta_o$. This figure indicates 
that normal-state properties
drive the spatially-dependent features of 
the superconductor's spectrum.\label{fig3}
}
\end{figure}

\noindent
were found for this potential. The curves showing the spectral
weight have been displaced from each other
so that they may be distinguished. Also
shown displaced in Fig.~3 is the spectral weight in the normal
state, normalized to the spectral weight in the homogeneous
metal.  {\it All} the quantities plotted in Fig.~3 are
{\it identical} to the accuracy of the calculation.
Fig.~3 is an illustration of a relationship between the
spectral weight in the normal state and the spectral weight
in the superconducting state, 
\begin{equation}
A({\bf r};\omega) = {A_{n}({\bf r})\over 2N_o}A_{sc}(\omega)\label{nonsum}
\end{equation}
where $A_n({\bf r})$ is the spectral weight in the inhomogeneous
{\it normal} state for energies near the Fermi surface and $A_{sc}(\omega)$
is the {\it homogeneous} superconductor's spectral weight as
a function of frequency. $2N_o$ is the normal state's
spectral weight far from the impurity. This expression is
valid for small $r$ and small $\omega$, the regime of
interest for STM on a superconductor. 
For $\omega$ of order $\Delta_o$, Eq.~(\ref{nonsum})
is valid for $r<\xi$.

We further illustrate the relationship of Eq.~(\ref{nonsum})
in Figure~4(a), which shows the LDOS for this
non-magnetic impurity as a function of voltage and position
calculated from Eq.~(\ref{STMLDOS}) with
$T={2\over 15}\Delta_o/k_B$. This temperature corresponds to 2K for
niobium.
There is no change in the energy {\it gap} due to this non-magnetic
potential. Figure~4(a) shows that it is merely the local
amplitude of the spectral weight which is reduced --- this
would manifest itself in a locally reduced oscillator
strength for an optical transition, or the reduction in the
tunneling current for an STM, which is directly proportional
to the LDOS 
}{
\noindent
shown in Figure~4(a). Figure~4(b) is an identical
calculation for a shorter coherence length, $\xi = 10k_F^{-1}$.
There appears to be little difference, although here a
localized state exists near the continuum. 

Figures~3 and~4(ab) show for the non-magnetic impurity that the
normal-state electronic structure determines 
the {\it spatial} dependence of the
superconductor's
$A({\bf r};\omega)$ for all frequencies including near the
energy gap. 
The potential strength of the impurity is orders of
magnitude greater than $\Delta_o$, and thus locally
mixes in states far

\begin{figure}
\epsfxsize=2.5in
\epsfysize=3.00in
\epsffile{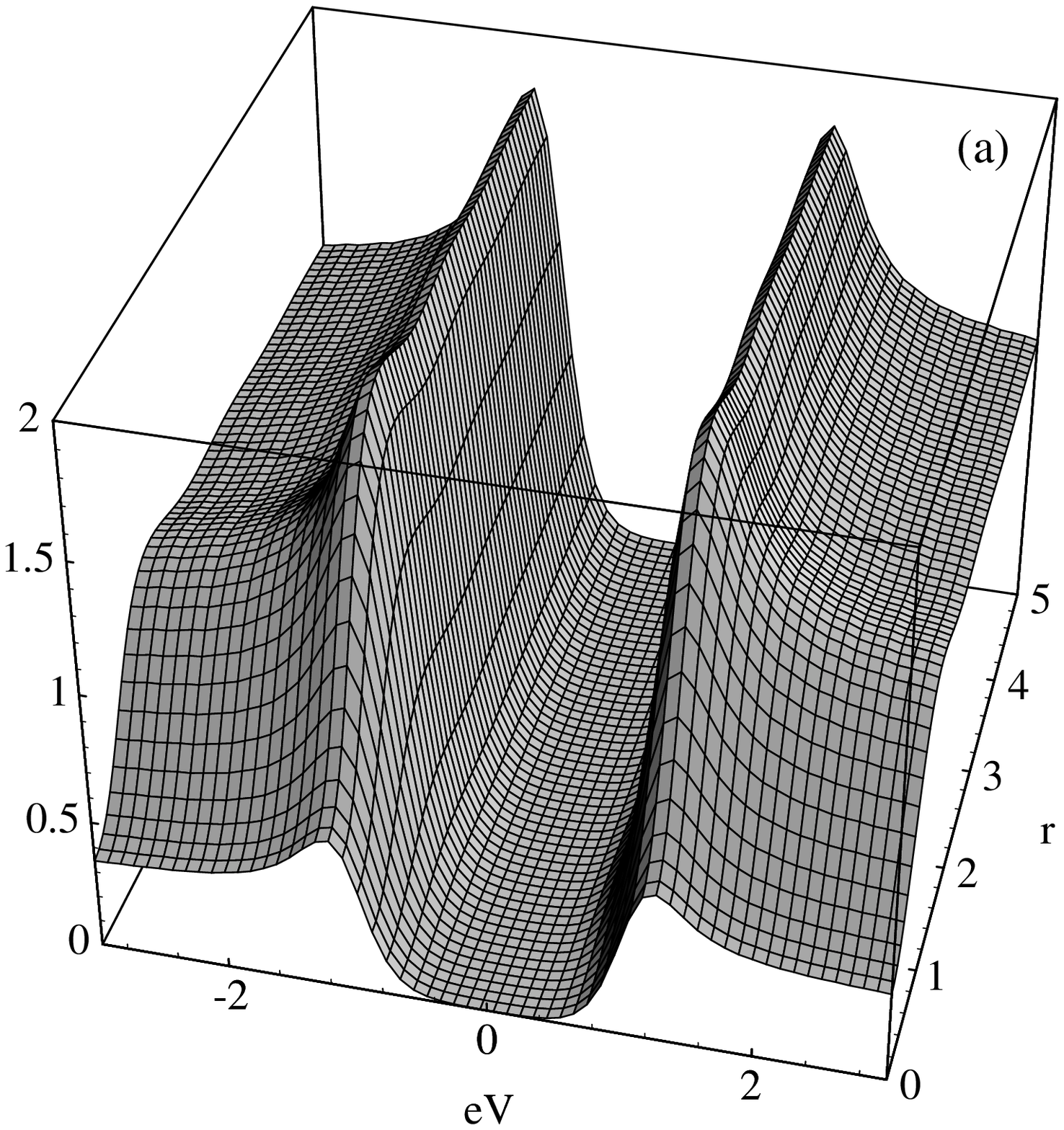}
\vskip\baselineskip
\epsfxsize=2.5in
\epsfysize=3.00in
\epsffile{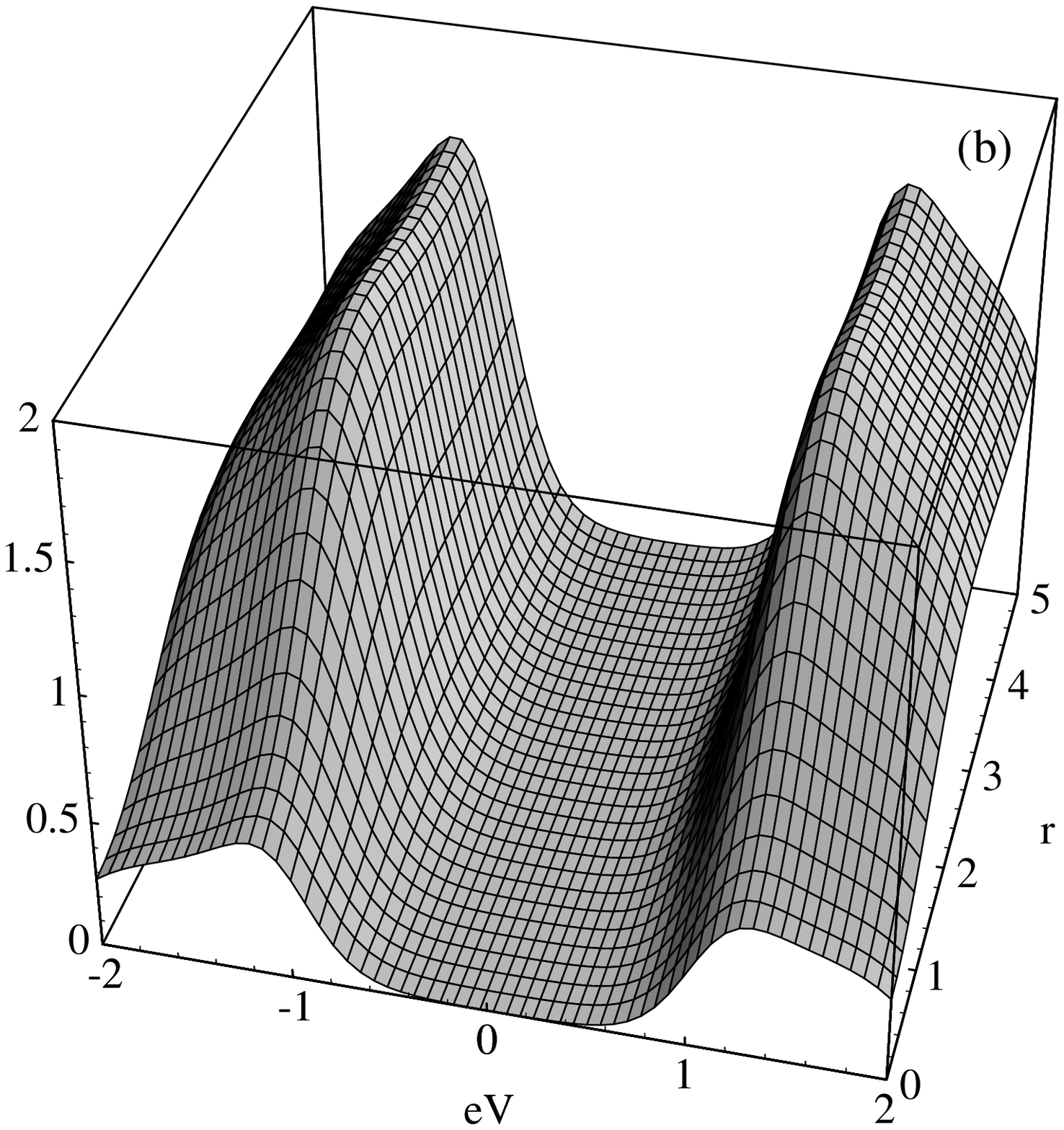}
\vskip\baselineskip
\caption{Differential conductance (LDOS) as a function of
voltage and position calculated around a non-magnetic 
impurity with $v_0 = 0.875$.
The spectrum is suppressed substantially in the vicinity of the 
impurity. The temperature is
$\Delta_o/7.5k_B$, 
which for niobium corresponds to about $2$K. (a) A coherence 
length appropriate for niobium, $\xi k_F = 449$. (b) A much shorter 
coherence length, $\xi k_F = 10$.  }
\end{figure}

}\narrowtext\twocolumn

\noindent
from the Fermi surface in the
homogeneous metal. These states are required
to construct probability densities which are
suppressed by 70\% near the impurity. The spatial structure
of the spectral weight in the normal state is essentially
identical to that seen in Fig.~3 over an energy range around
the Fermi surface which is orders of magnitude greater than
$\Delta_o$. 

Once the normal-state
band structure has been 
distorted by the presence of the
non-magnetic impurity, superconductivity is a small
perturbation within a narrower frequency range. The 
formation of the gap in the single-particle excitation
spectrum in the superconducting state is characterized by
the ``mixing''
of electron and hole amplitudes  to form quasiparticles
near the gap edge. These
quasiparticles, therefore, are constructed from single-particle
eigenstates of the metal which have already been strongly
distorted by the impurity potential.

Equation~(\ref{nonsum})
has important implications for spectroscopy on a
superconductor, for one of the procedures for normalizing
spectra taken at different lateral positions on a superconducting
surface is to assume that the
LDOS at a particular voltage much larger than $\Delta_o$ is
the same.
This is an attempt to
correct for possible changes in the tip-surface distance 
upon moving the tip laterally. A small change in the
tip-surface distance can have a strong effect on the
tunneling current.  Unfortunately this
procedure will prevent an experiment from seeing 
changes in the LDOS due to a non-magnetic impurity,
including the conductance oscillations described in
Ref.~\onlinecite{BFS}.

We now discuss the properties of the order parameter.
$\Delta({\bf r})$
is self-consistently determined, 
and is shown in Fig.~3 for small $r$ to be identical in spatial structure
to the normal-state spectral weight, 
\begin{equation}
{\Delta({\bf r})\over \Delta_o} = {A_n({\bf r})\over 2N_o}
={A({\bf r};\omega)\over A_{sc}(\omega)}.
\end{equation}
Since a non-magnetic 
potential repulsive to electrons attracts holes, and 
$\Delta({\bf r})$ depends equally on electron and hole
amplitudes, one might expect a non-magnetic potential to
have little effect on the spatial dependence of the order
parameter.  However the
allowable maximum spectral density of holes depends on the
spectral density of the electron band where the holes
reside, so if most electrons are excluded from the site,
holes will be effectively excluded as well. To emphasize
this point we note that the scaled anomalous spectral weight
Im$F({\bf r},{\bf
r};\omega)$ is identical to the scaled $A({\bf r};\omega)$ for the
frequencies shown in Fig.~3 (and for all relevant
frequencies for the self-consistency equation Eq.~(\ref{KSsc})).
Since Im$F({\bf r},{\bf r};\omega)$ is proportional to the
product of electron and hole amplitudes, and $A({\bf
r};\omega)$ is proportional to the electron amplitude
squared, the spatial structure of the electron and hole
spectral weights must be similar.  They are, since the
normal-state spectral weight $A_n({\bf
r};\omega)$ is roughly frequency independent around the
Fermi energy over an energy range
much greater than $\Delta_o$.

We now comment on the lack of localized states near the
non-magnetic impurity for $\xi k_F=449$ and the small
binding energy for the quasiparticle for $\xi k_F = 10$. The 
suppression of the order
parameter near the impurity may be considered to form an
attractive off-diagonal potential which may bind
quasiparticles. Localized states created by order-parameter
suppression
would be doubly degenerate, due to the two possible spin states
(see Eq.(\ref{allofit})).
Since the quasiparticle is half hole and
half electron, if the electron part is attracted and the
hole part is repelled, one might expect the effects of such
a non-magnetic potential on the quasiparticle to cancel. 
However the binding energy of the localized state is
an order of magnitude smaller ($\omega_o = (
1-2\times 10^{-4})\Delta_o$) than that found in Section IV.D for a
suppressed order parameter via pairing suppression. 
This 
may be explained by the well-known repulsive effect (quantum
reflection) of a
strong attractive potential on a quantum-mechanical
particle.
We find also that in the case of the
magnetic impurity that a large enough non-magnetic potential
of either sign will suppress the binding of a quasiparticle
to the impurity. 
We note here that the ratio of
the non-magnetic potential to the off-diagonal potential
($\delta\Delta({\bf r})$)
is much larger 
($v_0/N_o\Delta_o\sim 10^4$) 
than the ratios of the non-magnetic
potentials to the magnetic potentials considered below.

\subsection{Magnetic Impurity}

Recently, we have presented calculations of the LDOS (and
thus the differential conductivity in an STM experiment) in
the vicinity of a magnetic impurity\cite{Flatteshort}. These
calculations indicated that the spatial structure of the
electron amplitude of the localized state differed strongly
from the hole amplitude of the localized state. A further
result was that the spectrum should recover to
the homogeneous spectrum within a few atomic spacings.
Similarly
motivated calculations of the LDOS due to the $\ell=0$
localized state have been presented since then\cite{SalkBalat},
although these calculations did not address the continuum
LDOS. The two models used in Ref.\onlinecite{SalkBalat} were (1)
a $\delta$-function model solved using particle-hole symmetric Green's 
functions, but not self-consistently, and (2)
self-consistent calculations for a two-dimensional
tight-binding $s$-wave superconductor within the BdG
equations. 
The first method can only
model the normal-state properties properly for a
particle-hole symmetric band structure, such as at the Van
Hove singularity in a two-dimensional tight-binding band
structure. The
second method must contend with numerical finite-size effects,
which make it difficult to calculate
the continuum states.
A result obtained from the first method which is only true
for special band structures is that the spatial
structure of the electron and hole components of the
quasiparticle are the same.
The authors of
Ref.\onlinecite{SalkBalat} did raise the possibility of an
additional non-magnetic potential as a source

\break
\vskip\baselineskip
\vskip\baselineskip
\begin{figure}
\vskip\baselineskip
\vskip\baselineskip
\epsfxsize=2.5in
\epsfysize=2.0in
\epsffile{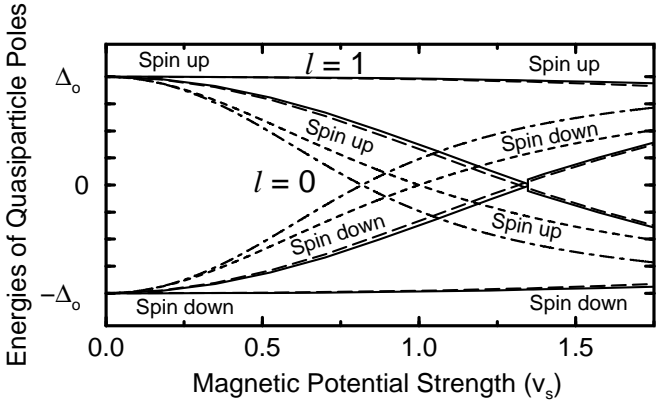}
\vskip\baselineskip
\caption{Energies of the quasiparticle poles as a function of the 
magnetic potential strength $v_s$
for the first two angular momentum states around the impurity. The 
spin-up and spin-down labels refer
to the band that the excitation resides in --- an excitation with 
negative energy (hole-like) in a
spin-up band is a spin-down hole. The solid line corresponds to 
$\xi k_F = 10$, the long-dashed
line to $\xi k_F = 449$, the short dashed line to the symmetric 
model of Shiba, and
the dot-dashed line to the result calculated with muffin-tin Green's 
functions and $\alpha=0.704$.
At a critical value of $v_s=v_{s0}^*$, the up poles cross to negative
energies and the down poles cross to positive
energies, indicating a change in the character of the ground
state. The kink evident in the solid and long-dashed lines is real, and due
to the discontinuous (at $T=0$) change in $\Delta({\bf x})$ at $v_{s0}^*$.
The unimportance of self-consistency can be gauged by the small 
difference between the short-coherence
length result and the long-coherence length result.
}
\end{figure}
\vskip\baselineskip
\vskip\baselineskip

\noindent 
of electron-hole amplitude asymmetry  in the spatial structure
of the localized state. We found\cite{Flatteshort} and will
explore below that there is, for realistic band structures,
electron-hole asymmetry without a
non-magnetic potential.

We will begin with a discussion of the energies of the
localized states around a magnetic impurity 
and the spin character
of those states. 
Solutions to 
Eq. (\ref{detm}) can be evaluated numerically. 
Figure~5 shows the dependence of the energies of the
localized 
state poles for the
first two angular momentum channels on the strength of the magnetic
potential. Results for a short coherence length ($\xi = 10k_F^{-1}$) 
are shown (solid line) as well as results for 
a long coherence length ($\xi = 449k_F^{-1}$).
The localized quasiparticle state for small $v_s$ is the spin state
attracted to the classical spin, which 
we will label up ($\uparrow$)\cite{FMAFM}.
As the potential strength increases, the excitation energy of each
angular momentum state decreases. At
some critical value $v_{s\ell}^*$
the localized state becomes a spin-down
excitation, the energy changes abruptly,
and then increases with increasing $v_s$.  
This behavior can be extracted from the analytic model 
(Eq.~(\ref{allofit})) as well.

Also shown in Figure~5 are the analytic results for the
pole energies for
$\alpha=0$ (Ref.~\onlinecite{Shiba}) and $\alpha=0.704$.  The
muffin-tin  model is no better than the
particle-hole symmetric model in predicting the localized
state energies. The muffin-tin model will prove more
successful at predicting the asymmetry between the electron
and hole amplitudes of the localized state.

The unimportance of self-consistency for determining the energies 
of the localized states can be gauged by the small difference between 
the short-coherence length result and the long-coherence length result.
The most important feature it determines in Fig.~5 is the size
of the discontinuity
in the localized state energy at $v_{s0}^*$. This discontinuity is
due to a discontinuous change in the order parameter at this magnetic
potential strength, a result pointed out in Ref.~\onlinecite{Flatteshort}
which will be discussed more below.

A magnetic impurity affects each member of a pair oppositely.
An up-spin electron is attracted to the impurity 
and a down-spin electron is
repelled.
The difference in energy between the two time-reversed electron
states leads to a locally suppressed pairing.
Therefore, the energy needed to break a
pair and create a localized
quasiparticle with angular momentum $\ell$
in the vicinity of the magnetic impurity, when 
$v_{s} < {v_{s\ell}}^*$, is reduced from
$2\Delta_o$ to $\Delta_o + \omega_\ell$.  One member of the
broken pair is a delocalized spin-down quasiparticle at the gap
edge (with energy $\Delta_o$).  The other member of the broken pair
is a {\it localized} spin-up quasiparticle with energy
$\omega_{\ell}$.  For $v_s>v_{s\ell}^*$ there is a spin-up
quasiparticle with angular momentum $\ell$ 
in the ground state of the superconductor\cite{Sakurai}.
An essential point about the ground state
of a superconductor containing classical magnetic impurities
is that at $T = 0$ when $v_{s} < {v_{s\ell}}^{\ast}$
(for all $\ell$) the ground state is composed entirely of paired
electrons.  When $v_{s} > {v_{s\ell}}^{\ast}$ for any
$\ell$ then the $T = 0$ ground state contains localized
quasiparticles as
well as pairs and there are new  low-energy excitations including
the re-formation of a pair as well as the excitation of a localized
quasiparticle to a higher energy localized or continuum state.

Evident from Eq.~(\ref{nambuG}) is that the
electron pole involves entirely single-particle
states within the spin-up band for $v_s<v_{s\ell}^*$, but
involves entirely states within the spin-down band for
$v_s>v_{s\ell}^*$.
The hole-like pole always involves single-particle states in the
opposite spin band from the electron-like pole. The source of the
quasiparticle amplitude for the various poles
is indicated in Fig.~5.

Figures~6(a), 6(b), and 6(c) show LDOS results for $v_s = 0.52$,
$0.875$, and $1.75$ respectively. They show the state
split off from the continuum, with a larger
electron-like amplitude than hole-like amplitude (6(a)), and then
lower in energy with an increased electron/hole asymmetry
(6(b)).  Finally the larger peak becomes hole-like (6(c)).
The asymmetry
between the electron-like and hole-like peaks becomes more
pronounced as $v_s$ increases. We note that the larger peak
is always associated with the spin-up band, whereas the
other is associated with the spin-down band. Despite the
apparent differences in peak size, the spatially integrated
electron spectral weight of the quasiparticle is equal to
the spatially inte-

\onecolumn
\widetext

\begin{figure}
\twomini{
\epsfxsize=20.5pc
\epsfysize=3.0in
\epsffile{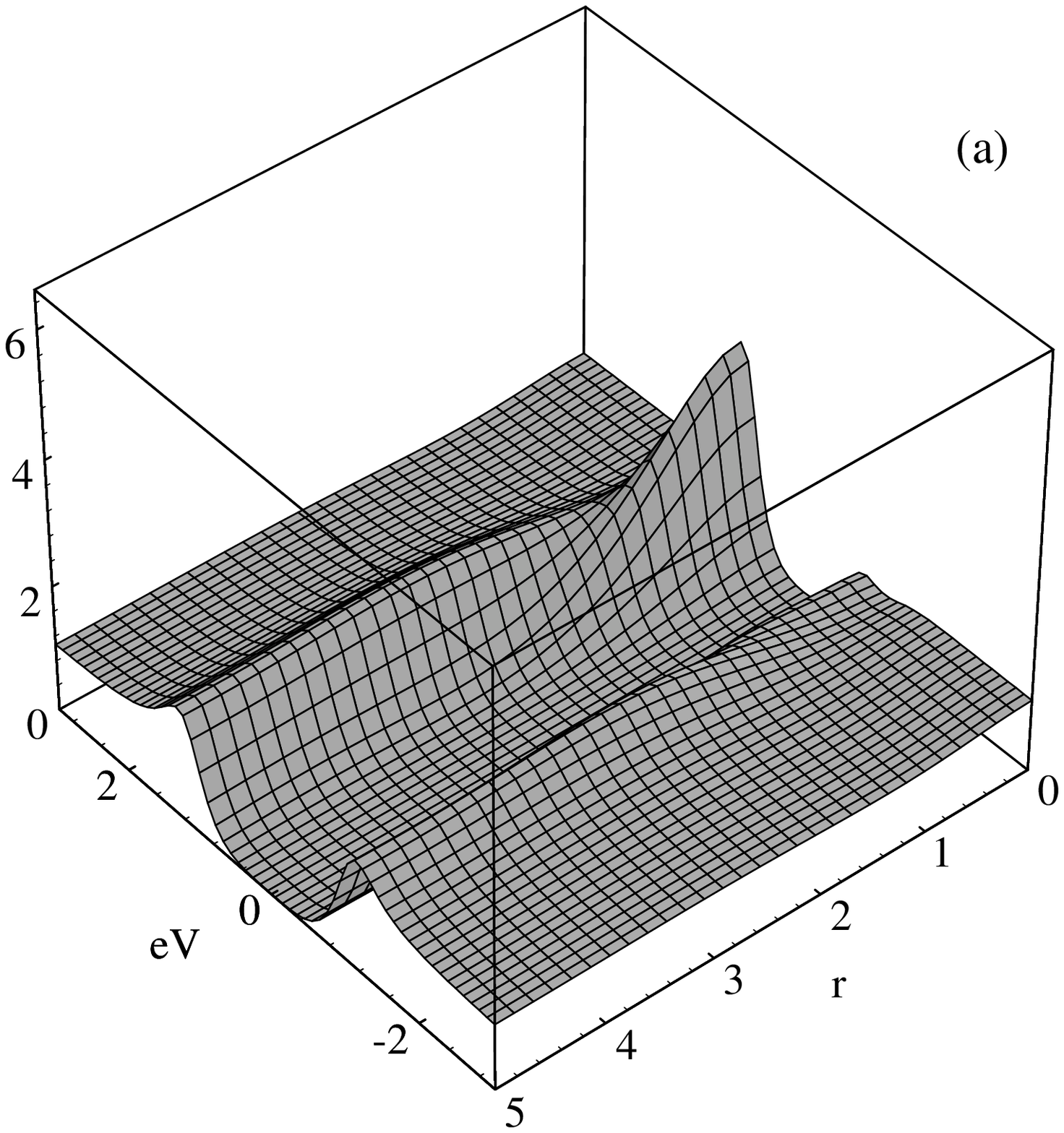}
\vskip\baselineskip
\epsfxsize=20.5pc
\epsfysize=3.0in
\epsffile{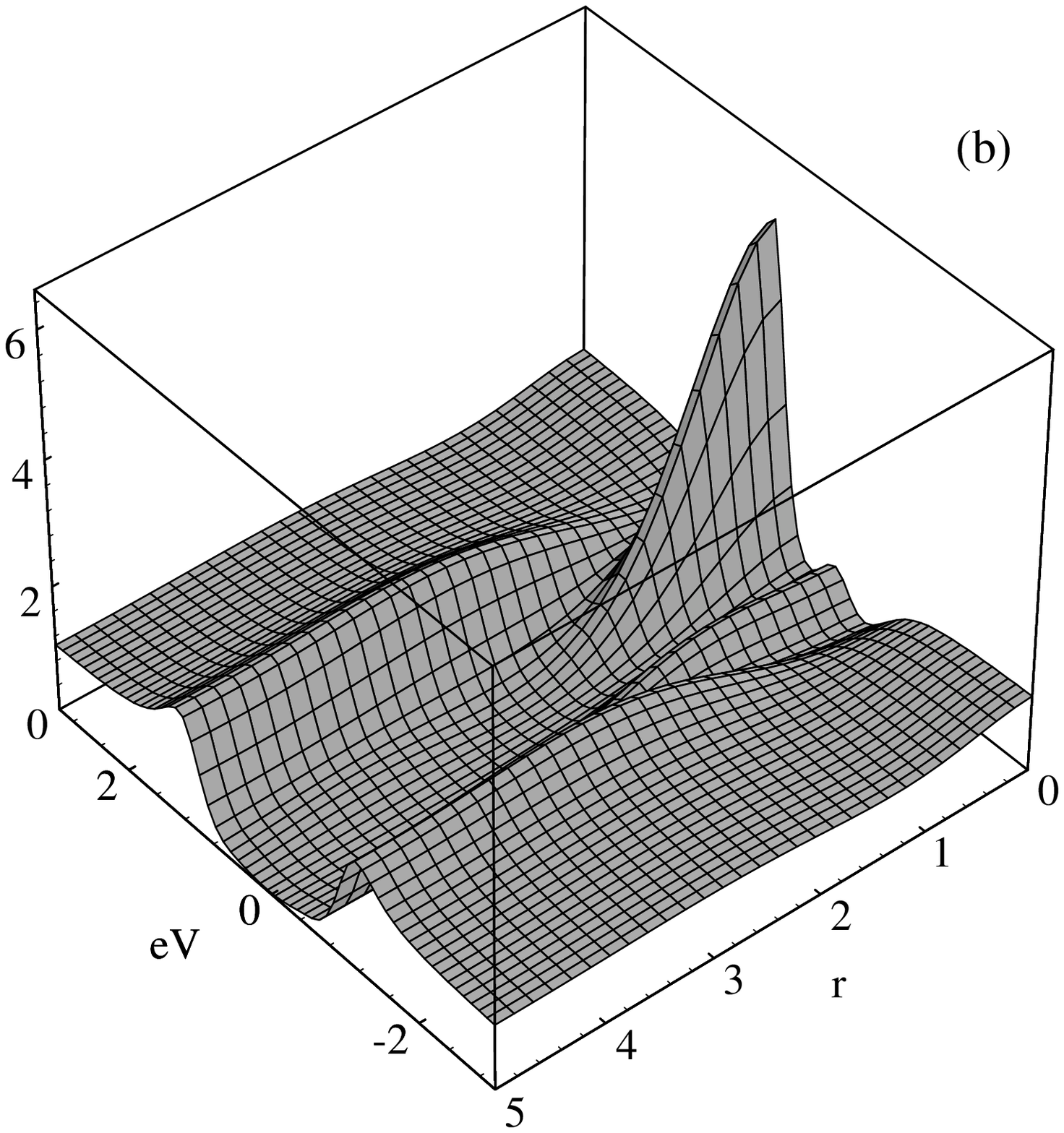}
}{
\epsfxsize=20.5pc
\epsfysize=3.0in
\epsffile{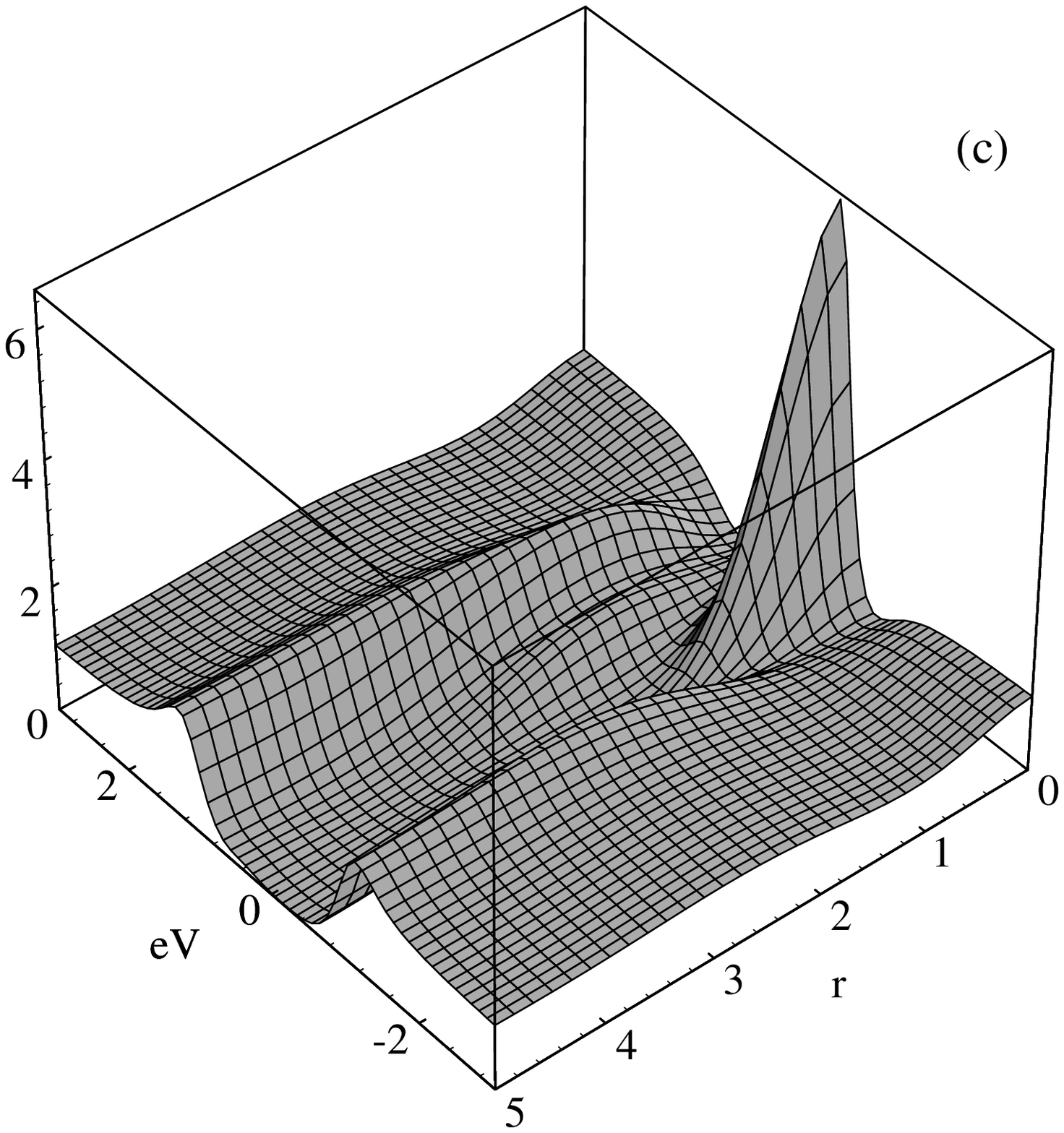}
\vskip\baselineskip
\epsfxsize=20.5pc
\epsfysize=3.0in
\epsffile{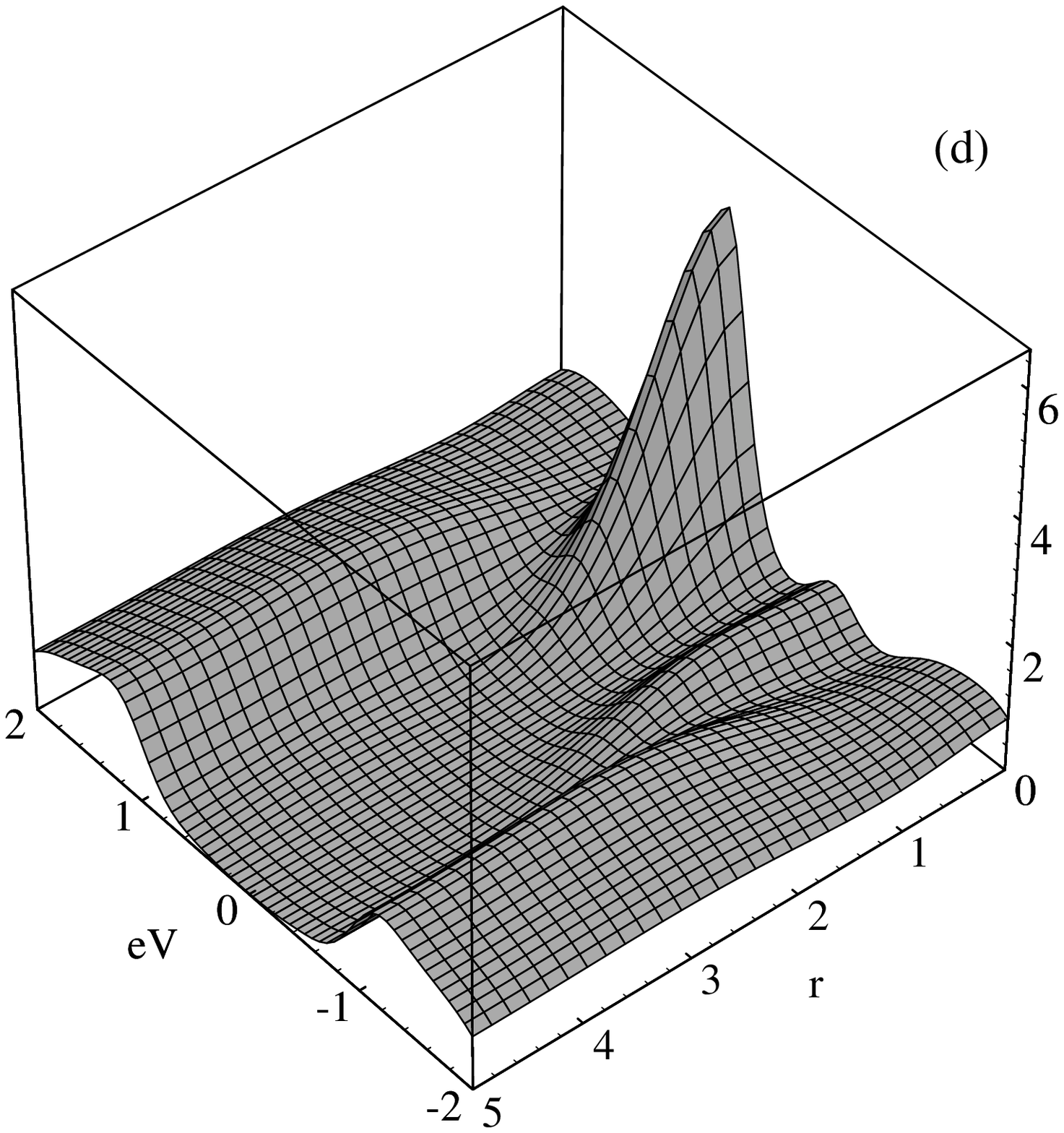}
}
\vskip\baselineskip
\caption{Differential conductance (LDOS) calculated around a 
magnetic impurity with (a) $v_s=0.5$, $\xi k_F =
449$, (b) $v_s=0.875$, $\xi k_F = 449$, (c) $v_s = 1.75$, 
$\xi k_F = 449$, (d) $v_s = 0.875$, $\xi k_F =
10$. All are calculated with $k_BT = \Delta_o/7.5$. In the 
progression from (a) to (c) the asymmetry
between the two peaks increases and the higher peak moves to lower 
energies, eventually becoming hole-like.  }
\end{figure}

\twomini{
\noindent
grated hole spectral weight. The
localized quasiparticle is always half electron and half hole for all
potentials examined here.
For $v_s<v_{s0}^*$ the spin-up band amplitude is
electron-like and the spin-down band amplitude is
hole-like.
At $v_{s0}^*$ (1.32 for niobium), the spin-up component 
becomes hole-like and the spin-down component becomes
electron-like, as required by the change in the spin of the
excitation.

Figure~6(d) shows the LDOS for a markedly different coherence
length, $\xi =10k_F^{-1}$, and $v_s = 0.875$. It is evaluated for
the same values of $\Delta_o/k_BT$ as Figs.~6(abc) and
looks almost
identical to Fig.~6(b). Since the localized state is broadened
by temperature through Eq.~(\ref{STMLDOS}), this 

}{
\noindent
is a manifestation of the proportionality of the spectral weight
to $N_o\Delta_o$ (Eq.~(\ref{MFSC})). 
Figure~7 shows the spectral weight at the
origin for the $\ell=0$ state and for the $\ell=1$ state at
its first maximum for $v_s = 0.875$ as a function of the inverse of the
coherence length, which 
is proportional to $N_o\Delta_o$. 
Figure~8 shows the spectral weight for $\xi = 449k_F^{-1}$
as a function of $v_s$ for the
spin-up and spin-down poles at the origin for the $\ell=0$
state and at the first maximum for the $\ell=1$ state. It is
clear that a non-magnetic potential is not necessary to
obtain an electron-hole asymmetry.

In Figure~9 we
show the asymmetry at the impurity as a function of $v_s$
for two values of $\xi$ --- a long coherence 

}\twocolumn\narrowtext

\begin{figure}
\epsfxsize=2.55in
\epsfysize=2.00in
\epsffile{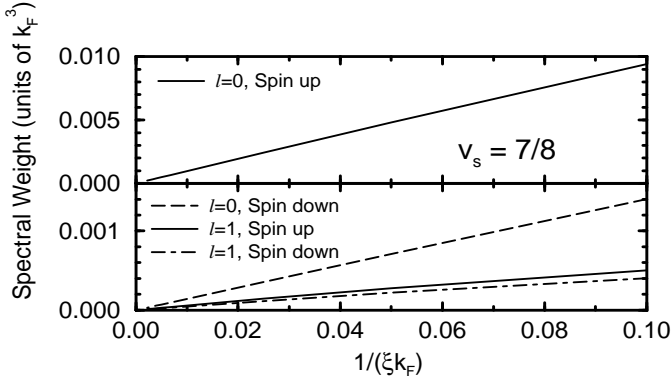}
\caption{Spectral weight at the impurity ($r=0$) for the first 
angular momentum channel, $\ell=0$, and
at the first maximum for the second angular momentum channel,
$\ell=1$, for poles in both the spin-up and spin-down bands, 
as a function of the inverse coherence length, showing a 
linear behavior.
The magnetic potential strength is $v_s = 0.875$.  }
\end{figure}

\noindent
length
appropriate for 
niobium, and a short coherence length. From 
Figure~7 it should
be evident that the asymmetry is not
sensitive to $\xi$. It is, however, predicted extremely well
by the normal-state spin-up and spin-down band spectral
weight asymmetry at the 
impurity (also shown in Figure~9).
We can therefore 
conclude that as
with the non-magnetic impurity (Figure~4), the spatial
structure of the spectral weight is a normal-state property.
We further show in Figure~10 for $v_s = 0.875$ 
the $\ell=0$, spin-up band and spin-down 
band projections of the normal-state spectral weight to compare
with the localized state spin-up band and spin-down band
spectral weights for two values of $\xi k_F$. 
The normal-state and long-coherence length calculation
are practically indistinguishable. 
The insets show
$r^2A(r)$, which removes the rapid power-law decay of the
state. The localized states for all angular momenta $\ell$ will
decay as the power law $r^{-2}$. For the short-coherence length 
calculation the

\begin{figure}
\epsfxsize=2.55in
\epsfysize=2.00in
\epsffile{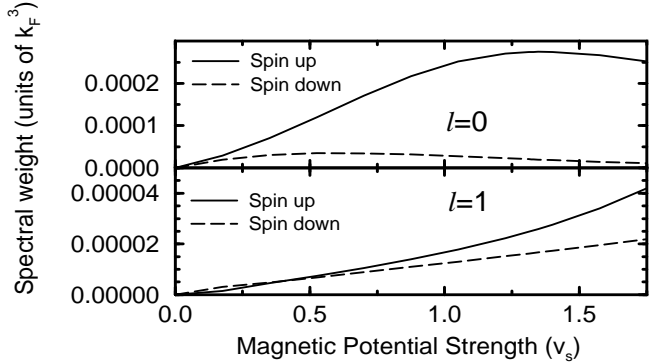}
\caption{Spectral weight at the impurity ($r=0$) for the $\ell=0$ 
channel for $\xi k_F = 449$ as a function
of magnetic potential strength for poles in both the spin-up and 
spin-down bands. The spectral weight in the spin-up band
pole of the $\ell=0$ localized state saturates at large $v_s$.
Also shown are the spectral weights at the first maximum for 
the $\ell=1$ localized states.  }
\end{figure}

\begin{figure}
\epsfxsize=2.50in
\epsfysize=2.000in
\epsffile{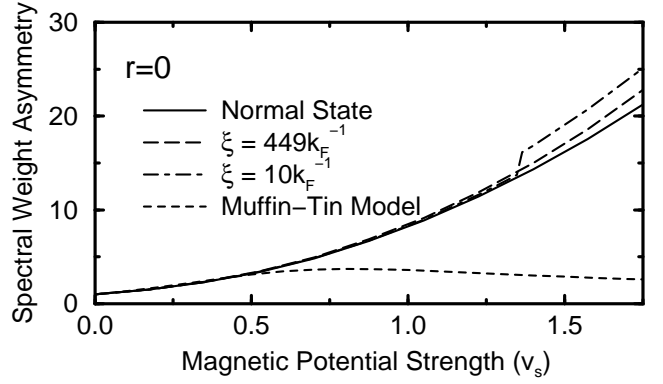}
\caption{Ratio of the spectral weight in the spin-up band and in the 
spin-down band at the
impurity ($r=0$) as a 
function of magnetic potential strength. This is plotted for the 
normal state $\ell=0$ projected
Green's functions (solid line) as well as for the localized states
for niobium ($\xi k_F = 449$,
long dashed line), for $\xi k_F = 10$ (dot-dashed line), and for the 
muffin-tin model. The muffin-tin
model is only successful for $v_s<0.5$, but that is due to a 
breakdown in describing the normal state.
The normal-state electronic structure is a good predictor of the 
superconductor's electronic
structure for the entire range of $v_s$.  }
\end{figure}
\noindent
effect of an exponential envelope is also visible.
In the analytic result the
exponential envelope should have a range 
$R = \pi\xi/2\sqrt{1-(\omega_o/\Delta_o)^2}$, which corrects to
better than 1\% the discrepancy in Figure~10.
The power-law 
\vskip1.5pc

\begin{figure}
\epsfxsize=2.50in
\epsfysize=2.00in
\epsffile{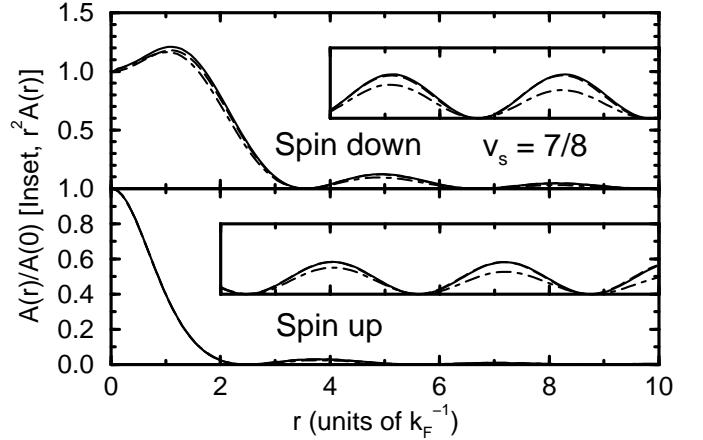}
\vskip1.pc
\caption{Spectral weights for $\ell = 0$ localized state in the up and down
bands for $v_s = 0.875$. The solid line is the normal state $\ell=0$ 
projected spectral weight, 
the long-dashed
line is the localized state in a superconductor with $\xi k_F =449$, 
and the 
dot-dashed line is for $\xi k_F = 10$.
The inset shows the spectral weight multiplied by $r^2$ to remove the
algebraic decay. The normal-state and long-coherence length results 
are practically
indistinguishable. The deviation shown in the short-coherence length 
superconductor's
spectral weight is fit to within 1\% by the exponential decay factor 
described
in the text.  Hence the spatial structure of the spectral
weight of the superconductor's localized state is well-predicted by 
the normal state
spectral weight.
 }
\end{figure}
\noindent 
fall-off and exponential envelope  
can be seen directly 
from Eq.~(\ref{bareg}) and
Eq.~(\ref{MFSC}).

We can summarize these comments with a general equation,
similar in concept to that for the non-magnetic impurity,
Eq.~(\ref{nonsum}).  That is, for a localized quasiparticle state of
spin $\sigma'$, the spectral weight of a localized state
with angular momentum $\ell$ would be

\begin{equation}
A_\sigma({\bf r};\omega_{\ell}) = B A_{n\sigma}({\bf r},\ell)
{\rm e}^{
-\left({2r\over \pi\xi}\right)
\sqrt{1-\left({\omega_\ell\over\Delta_o}\right)^2}
}\delta(\omega
-\sigma\sigma'\omega_\ell),\label{magsum}
\end{equation}
where $B$ is a normalization factor so that the spectral
weight of the state integrates to one, and $A_{n\sigma}({\bf
r},\ell)$ is the angular-momentum $\ell$ projection of the
{\it normal}-state spectral weight in the spin $\sigma$ band.
We note that for small $r$ there is an approximate relationship between 
the superconducting
state's spectral weight and the normal state's spectral
weight in each spin band,
\begin{equation}
{1\over 2E} \int_{-E}^{E} d\omega 
A_\sigma ({\bf r};\omega) = A_{n\sigma}({\bf r}),
\label{sumrule}
\end{equation}
where $\Delta_o \ll E \ll \epsilon_F$. This, in connection
with Eq.~(\ref{magsum}), implies a dependence on the
normal-state structure of the 
continuum spectral weight around the magnetic impurity.

We now return to the structure of $\Delta({\bf x})$. 
This quantity, which is not directly observable, formed the
focus of several investigations of the local structure around
a

\begin{figure}
\epsfxsize=2.5in
\epsfysize=2.00in
\epsffile{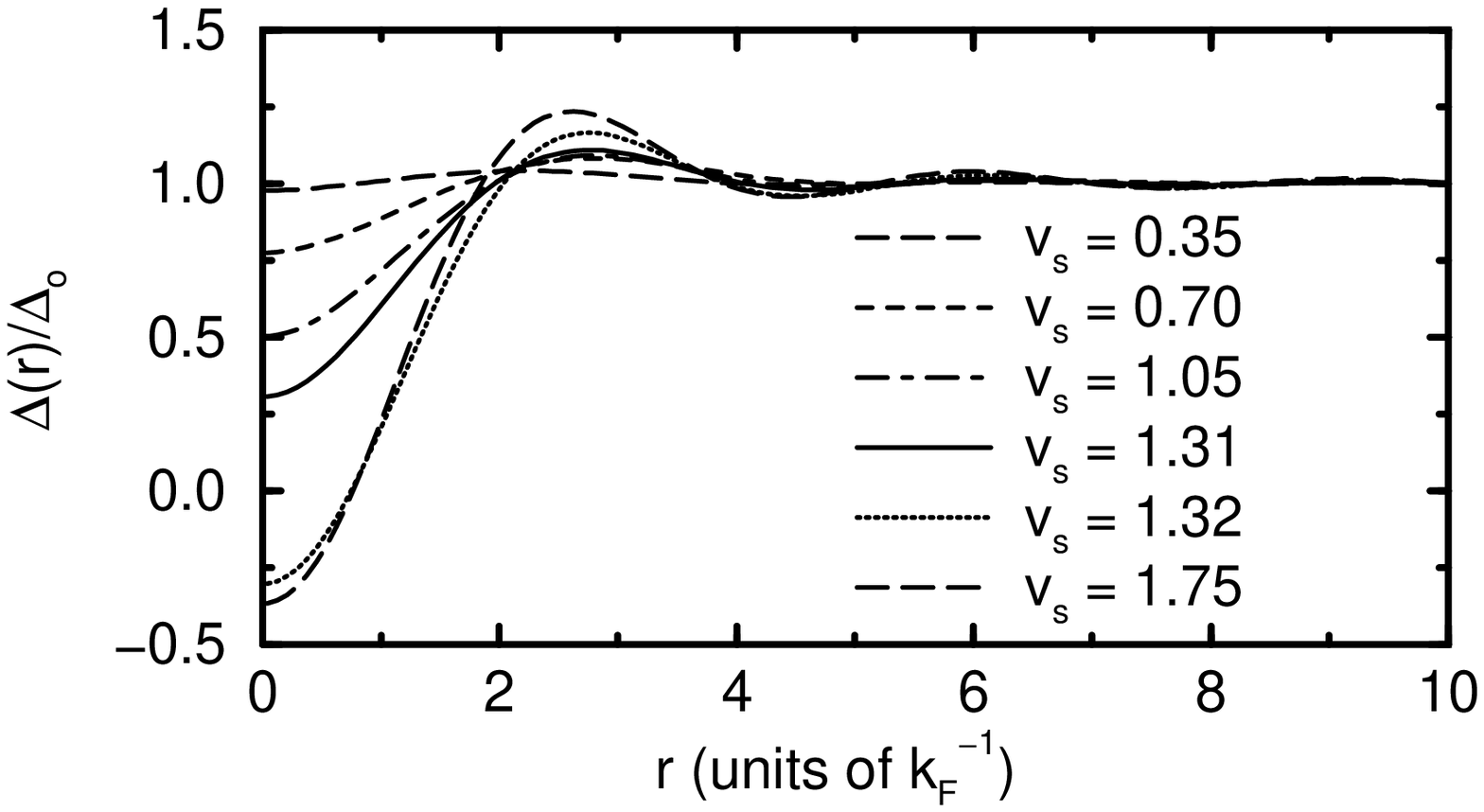}
\epsfxsize=2.5in
\epsfysize=2.00in
\epsffile{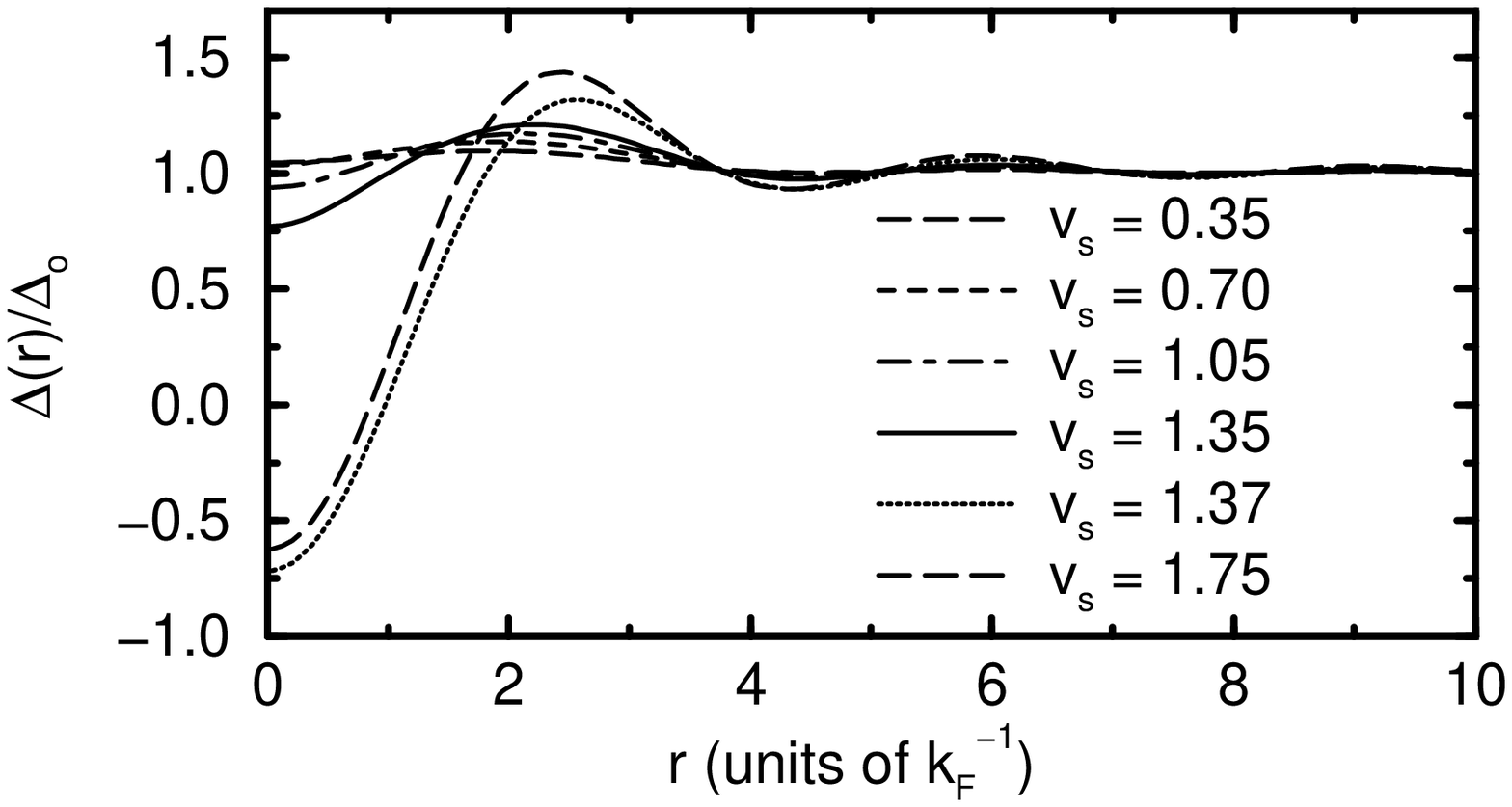}
\caption{Order parameters as a function of distance from the impurity $r$
calculated for several magnetic potential strengths for (a)
$\xi k_F = 449$ and (b) $\xi k_F = 10$. In both cases there is a 
discontinuous change
in the order parameter when $v_s$ passes through the critical strength
 $v_{s0}^*$.  }
\end{figure}

\begin{figure}
\epsfxsize=2.5in
\epsfysize=2.00in
\epsffile{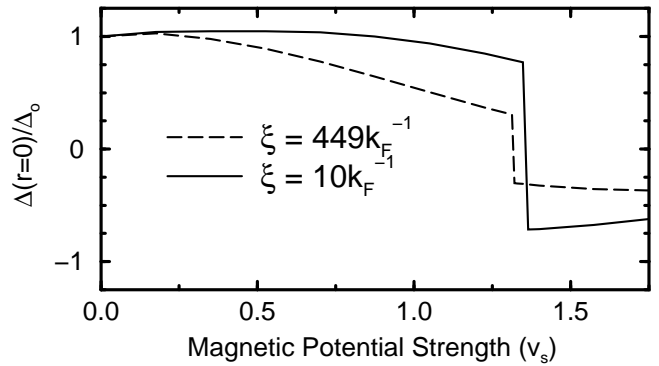}
\caption{Order parameters at the impurity ($r=0$) as a function of 
magnetic potential
strength $v_s$ for two values of the coherence length. The 
discontinuity in the order
parameter at $v_{s0}^*$ is much larger for the short-coherence-length 
superconductor.}
\end{figure}

\noindent
magnetic impurity.
The oscillation of the order
parameter around a magnetic impurity was first evaluated without
self-consistency\cite{Tsuzuki,Heinrichs,Kummel}.
A self-consistent calculation of 
the order parameter at the impurity and very far away
for {\it weak} impurity potentials was done by
Schlottmann\cite{Schlottmann}.

As
shown in Figure~1, for large values of $v_s$, $\Delta({\bf
x}={\bf 0})<0$. Sign changes in $\Delta$, as seen in pair
tunneling,  have been suggested for magnetic impurities in the
barriers of Josephson junctions\cite{junc1,junc2,junc3}.
The sign change in $\Delta({\bf 0})$ occurs (at
$T=0$) precisely at $v_{s0}^*$. 
Due to the spin and frequency symmetries of Eqs.~(\ref{KS}-\ref{potn}),
the anomalous
spectral weight ${\rm Im}F ({\bf r},{\bf r}, \omega)$
associated with the spin-up pole is always equal and
opposite to the anomalous spectral weight associated with
the spin-down pole. 
As the
pole in the spin-up band goes from electron-like
($\omega>0$) to
hole-like ($\omega<0$) and the pole in the spin-down band
goes from
hole-like to electron-like the contribution to $\Delta({\bf
0})$ changes sign abruptly at $T=0$. The
$\Delta({\bf r})$ resulting from several values of
$v_s$ and two values of the coherence length are shown in Figure~11(ab).
The discontinuity at $v_{s0}^*$ is more pronounced for
shorter coherence lengths since the localized state's
spectral weight is more concentrated at the impurity
(Eq.~(\ref{magsum})).
$\Delta({\bf 0})$
as a function of $v_s$ is shown in Figure~12 for two values
of the coherence length.
At $T>0$ the
transition would be smoothed somewhat.

The behavior of $\Delta({\bf 0})$ as a function of $v_s$
comes
from the introduction at $v_{s0}^*$ of a quasiparticle into
the ground
state of the system.
The spin-up quasiparticle localized near
the impurity in the ground state suppresses the local order
parameter. For time-reversal invariant 
potentials one
cannot make $\Delta({\bf r})$ negative by inserting
a single 
quasiparticle, since
the suppression from one quasiparticle is cancelled by the
lack of suppression from its unexcited Kramers doublet
partner. For a spin-dependent potential, however, the anomalous
spectral weight near the impurity may be almost entirely
contributed by the single low-energy localized state. When
a quasiparticle is present in the ground state,
the ground state  has spin ${1\over 2}$ up\cite{Sakurai,SalkBalat}
and a negative $\Delta({\bf r})$\cite{Flatteshort}.
Exciting the low-energy state for $v_s>v_{s0}^*$ 
removes the spin-up quasiparticle, and therefore 
{\it increases} $\Delta({\bf 0})$,
whereas excitation of quasiparticles typically reduces
$\Delta({\bf x})$ (which is the case for $v_s<v_{s0}^*$).
Also, exciting the low-energy state {\it reduces} the
induced spin of the superconductor at the impurity.

\subsection{Combined Magnetic and Non-Magnetic Potentials}

We now discuss the addition of a non-magnetic potential to
the magnetic potential.
It had been suggested\cite{SalkBalat} that
introducing a $v_0$ with a $v_s$ would provide electron-hole
asymmetry. We
find that it does change the asymmetry, which we show in
Figure~13 for a particular $v_s$, but that once again this
is a normal-state property. 
 The relationship between the
normal-state spectral weights and the superconducting-state
spectral weights of Eq.~(\ref{magsum}) still holds.
Introducing $v_0$ also alters
the localized-state energies (see Eq.~(\ref{allofit})),
which we show in Figure~14 for $v_s = 0.875$ and $\xi k_F=449$.
The presence of a non-magnetic potential may affect the
value of $v_{s0}^*$\cite{SalkBalat}. We show in Figure~15 a
partial diagram of the ground state as a function of the parameters $v_s$
and $v_0$ for $\xi k_F = 449$. We note that 
the boundary between the two ground
states is
not shifted much from $\xi k_F=449$ to $\xi k_F=10$, hence the
condensation energy is not very significant in determining
this boundary. This fact, and the observation that the
energy of the localized state (in the absence of
self-consistency) relative to the homogeneous
order parameter does not depend on any parameter of the
superconducting state (Eq.~(\ref{allofit})) suggests that an
understanding of the transition at $v_{s0}^*$ may be
assisted by normal-state 
concepts. We hope to address
this issue in a future 

\vskip\baselineskip

\begin{figure}
\epsfxsize=2.5in
\epsfysize=2.00in
\epsffile{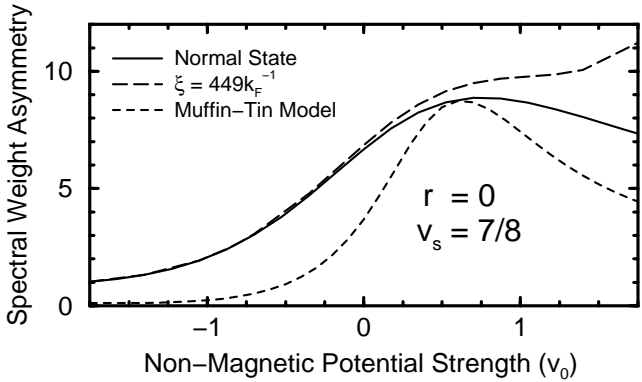}
\caption{Ratio of the spectral weight at the impurity in the spin-up 
band to the spin-down band
for $v_s = 0.875$ and $\xi k_F = 449$
as the non-magnetic potential $v_0$ varies. In a 
similar result to that seen
in Fig.~9, the normal-state spectral weights are good predictors of 
the superconducting state's
spectral weight. We note that the curves are not symmetrical around 
$v_0 = 0$, which results from
a realistic band structure without particle-hole symmetry.
}
\end{figure}

\begin{figure}
\epsfxsize=2.5in
\epsfysize=2.00in
\epsffile{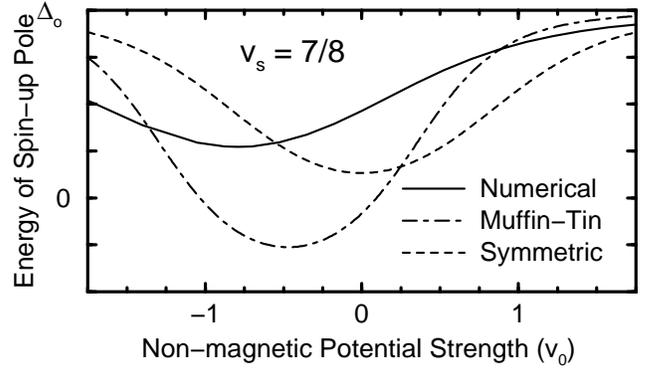}
\caption{The energy of the spin-up pole is shown as a function of 
non-magnetic potential
strength for $v_s = 0.875$ and $\xi k_F = 449$. The 
energy of the spin-down pole is just 
the negative of the energy
of the spin-up pole. The analytic models do not perform particularly 
well in reproducing the pole
energies, although the muffin-tin model does show a similar 
asymmetry around $v_0=0$ to the numerical calculations.  }
\end{figure}

\noindent
publication.

We now mention an example designed to
showcase the drawbacks of Ginzburg-Landau theory. We present
in Figure~16 the order parameter for a mixed magnetic and
non-magnetic impurity ($v_s = 7/8$, $v_0 = -7/4$, and $\xi
k_F = 449$). The order
parameter is everywhere larger than in the homogeneous
superconductor, however, the presence of a localized state
within the gap indicates that superconductivity has been
weakened around the impurity. 
Since Ginzburg-Landau
theory focuses on the 
order parameter and ignores the
quasiparticle structure, 
a Ginzburg-Landau perspective would
incorrectly predict an enhancement of superconductivity in the region.

We conclude by attempting to provide some guidance for attempts to extract
impurity potentials from STM 
measurements. Figure~17 shows
{\it normal-state} $dI/dV$'s for various potentials. These
curves should also represent 

\begin{figure}
\epsfxsize=2.5in
\epsfysize=2.0in
\epsffile{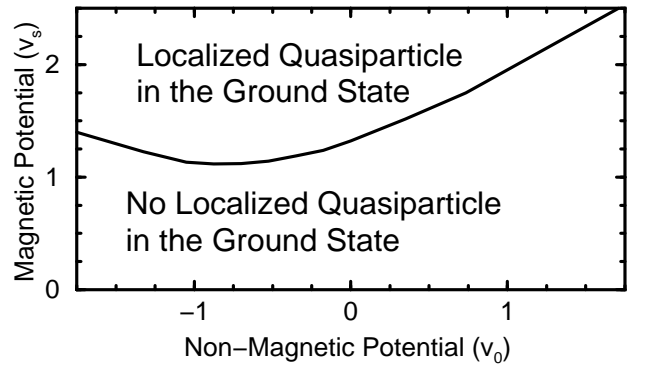}
\caption{Calculated boundary between two ground states around the 
magnetic impurity for $\xi k_F = 449$. For a large
enough magnetic impurity strength a quasiparticle is bound in the 
ground state. The minimum 
magnetic impurity strength depends on the non-magnetic impurity 
strength. For still larger magnetic
impurity strengths there would be ground states with more than one 
bound quasiparticle.
}
\end{figure}

\begin{figure}
\epsfxsize=2.25in
\epsfysize=.975in
\epsffile{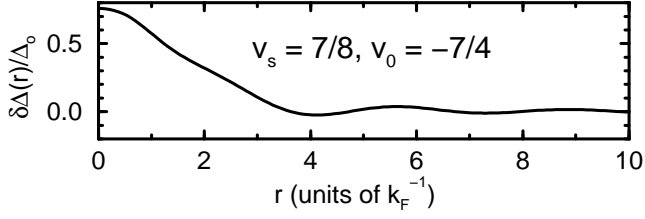}
\vskip\baselineskip
\vskip\baselineskip
\caption{The change in order parameter around an attractive 
non-magnetic potential combined with a
magnetic potential for $\xi k_F = 449$. The 
order parameter is larger at the impurity 
than in the homogeneous
superconductor, yet there exists a localized state in the gap.
}
\end{figure}

\noindent
the frequency-averaged spectral
weight measured in the superconducting state (see
Eq.~(\ref{sumrule}))\cite{usefulQ}.

The enhancement or suppression of spectral weight near the
origin is particularly sensitive to $v_0$.  A measurement of
this quantity, the energy of the localized state and the
asymmetry of the electron and hole amplitudes at the
impurity overconstrains $v_0$ and $v_s$, given an assumption
of the shape of the potential. To extract information about
the potential's detailed shape would require a fitting
procedure using the differential conductivity at various
positions. If, for some reason, the spin-down amplitude were
too small to measure, it may remain possible to constrain the
potential strength using the frequency-integrated spectral
weight and the localized-state energy. 

It seems appropriate to mention again the tendency to normalize
spectra according to the LDOS measured at energies much larger
than $\Delta_o$. Since the normal-state LDOS near the impurity
changes substantially
in the presence of magnetic or non-magnetic impurities, 
an experiment performed using such a normalization procedure 
would yield impurity parameters of questionable validity.

\subsection{Pairing Suppression}

The pairing potential, $\gamma({\bf x})$ in Eq.~(\ref{KSsc}),
may also have spatial structure. When this parameter is
changed it induces a change in the order parameter which
produces an off-diagonal potential felt by the
quasiparticles. We set $v_s=v_0=0$ so that there is no magnetic or
non-magnetic potential to compete with the order parameter
change. Figure~18 shows the order
parameter around a short-range suppression,
\begin{equation}
\gamma({\bf x}) = \left[1-{\rm e}^{-(k_Fr)^2}\right]\gamma_o,
\end{equation}
for two values of the coherence length. The order
parameter is strongly suppressed and since $\gamma({\bf 0})=0$, 
$\Delta({\bf 0})=0$. For long
coherence lengths, this change in the order parameter has no
effect on the local density of states (shown in Figure~19(a)).
It is possible to localize quasiparticle states, however, at
shorter coherence lengths. These 

\begin{figure}
\epsfxsize=2.5in
\epsfysize=4.2in
\epsffile{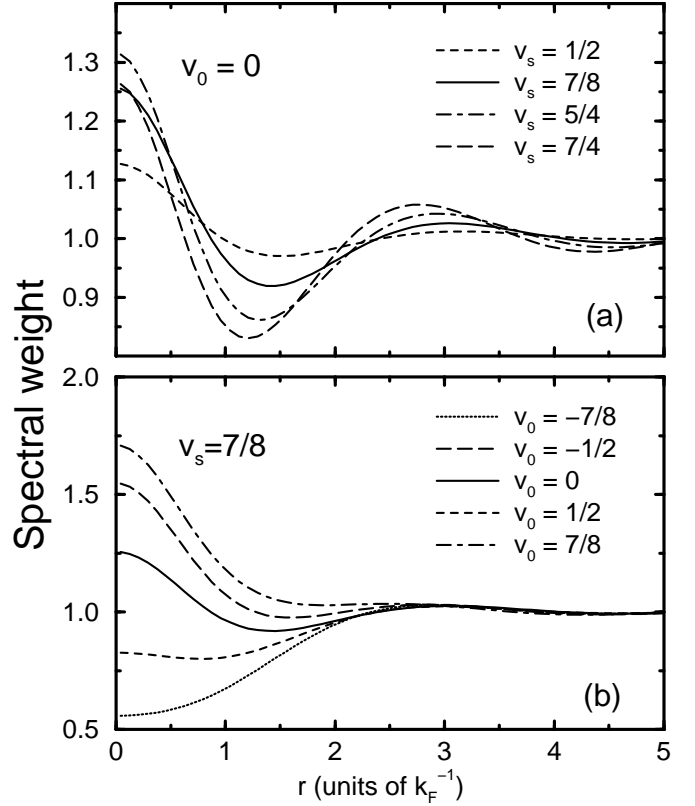}
\vskip\baselineskip
\caption{Spectral weights in the normal state as functions
of the distance from the impurity for several 
combinations of magnetic and non-magnetic
potentials (a) only a magnetic potential, and (b) a fixed magnetic 
potential $v_s = 0.875$ and a varying non-magnetic
potential. By making measurements around the impurity in the normal 
state (or integrating the 
superconductor's spectrum over a frequency much larger 
than $\Delta_o$), information about the
structure of the impurity may be obtained.
}
\end{figure}

\noindent
can produce features in the
local density of states which are visible. One such case is
shown in Figure~19(b). The energy of the localized state is
$\omega_o = (1-4\times 10^{-3})\Delta_o$. Whereas a 
non-magnetic potential changes
the local 

\begin{figure}
\epsfxsize=2.5in
\epsfysize=2.00in
\epsffile{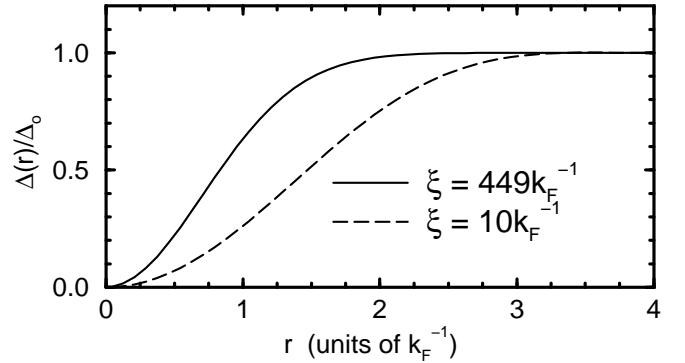}
\caption{Order parameter as a function of distance $r$ from
a defect with a suppressed pair potential, but no single-particle
potential. 
}
\end{figure}

\onecolumn\widetext

\begin{figure}
\twomini{
\epsfxsize=20.5pc
\epsfysize=3.0in
\epsffile{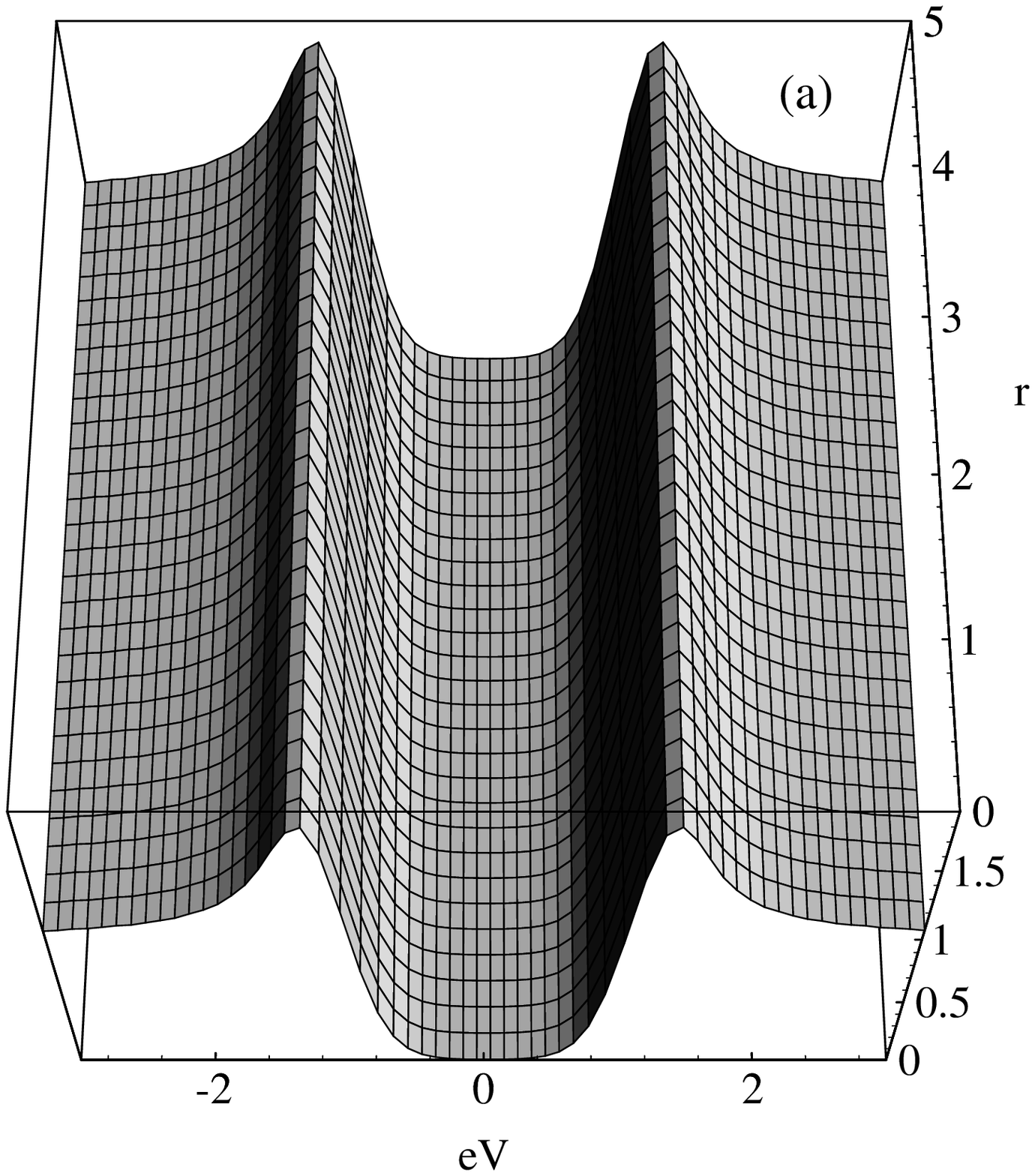}
}{
\epsfxsize=20.5pc
\epsfysize=3.0in
\epsffile{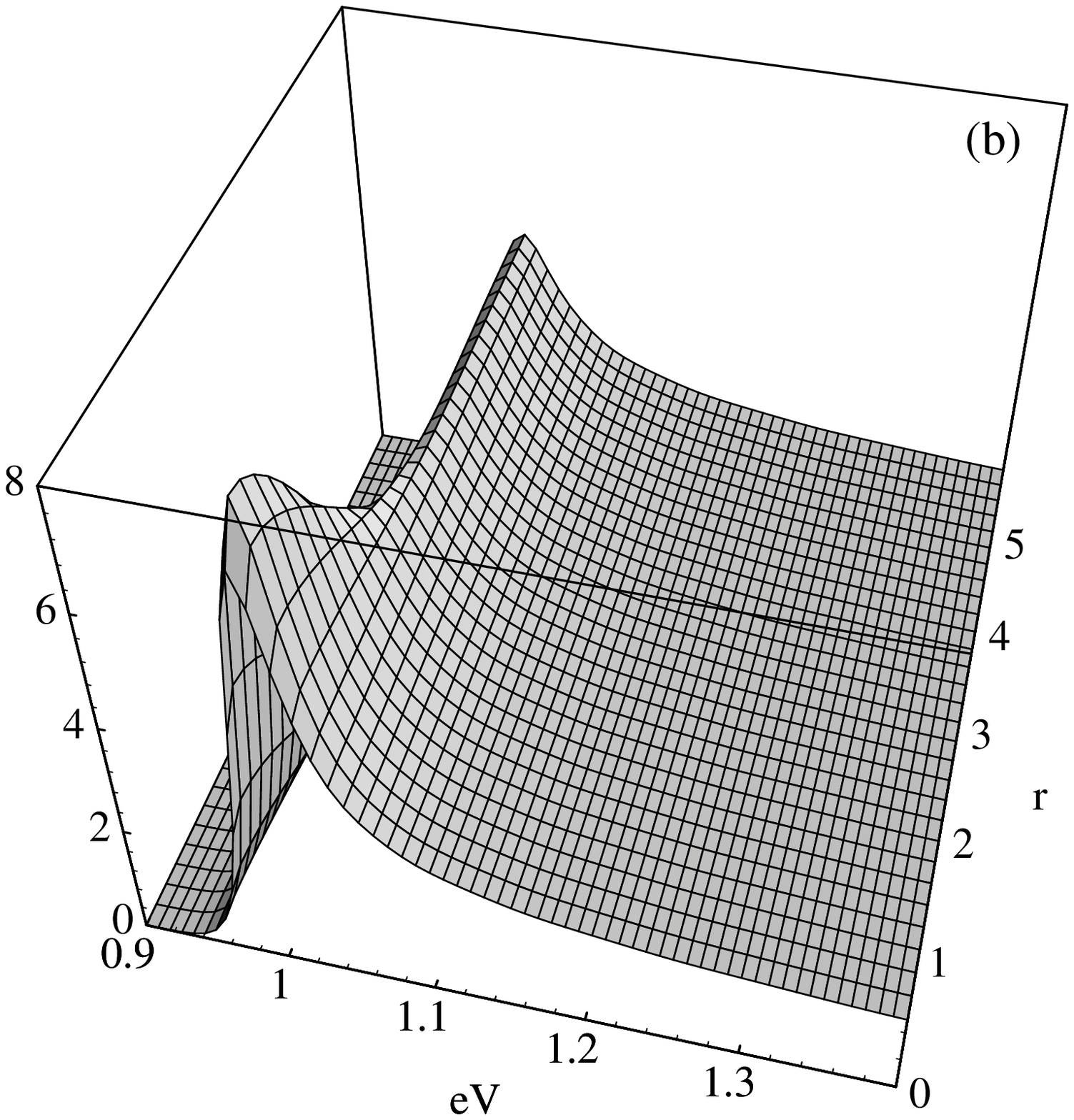}
}
\caption{Differential conductance (LDOS) around a defect with a 
suppressed pair
potential, but no single-particle potential. (a) $\xi k_F = 449$, 
$k_BT = \Delta_o/7.5$ ---
there is no evidence of any
change in the spectrum due to the order parameter suppression shown 
in Fig.~18. (b) $\xi k_F = 10$, 
$k_BT = \Delta_o/100$ --- a localized state very near the continuum 
enhances the continuum edge seen in
tunneling near the defect.
}
\end{figure}

\noindent 
density of states without significantly
changing the energy gap,
a pairing suppression has a very weak effect on both,
especially in the long-coherence length limit.

\section{Strong-Coupling and Anisotropic Order Parameters}

A few observations are in order concerning the extension of
this formalism to systems where the homogeneous
order parameter has important frequency or spatial structure.
The Gor'kov equation
(Eq.~(\ref{KS})) changes due to the 
more general form for the off-diagonal
potential originating from the order parameter. Taking this
opportunity to generalize,

\widetext
\begin{equation}
\int d{\bf x''}d{\bf x'''}
\left[ {\bf \delta}({\bf x}-{\bf x''}){\delta({\bf x''}-{\bf
x'''}) - 
{\bf g}({\bf x},{\bf x''}; \omega){\bf V}({\bf 
x''},{\bf x'''}
;\omega})\right] {\bf G}({\bf x'''},{\bf x'};\omega) = {\bf 
g}({\bf x},{\bf x'};\omega)\label{KSgen}
\end{equation}
where 
\begin{equation}
{\bf V}({\bf x''},{\bf x'''};\omega) = \left( 
\begin{array}{cc}
V_{e\uparrow}({\bf x''};\omega)\delta({\bf x''}-{\bf x'''})
&\delta\Delta({\bf x'',x''';\omega})\\ 
\delta\Delta({\bf x'',x''';\omega})& 
V_{h\uparrow}({\bf x''};\omega)\delta({\bf x''}-{\bf x'''})
\end{array}
\right).
\end{equation}
\vskip0.5pc
\twomini{
\vskip0.3pc
\noindent The diagonal terms are general potentials, possibly
frequency-dependent, effective on spin-up electrons $(V_{e\uparrow})$ 
and spin-up holes $(V_{h\uparrow})$ for $\omega>0$.
Since this potential is diagonal in
frequency, as is the Gor'kov equation, 
the frequency structure of the order
parameter does not add any additional complication to
numerically solving the Gor'kov equation. However,
the addition of spatial structure
to the pairing has added another integral over the volume
to the Gor'kov equation, and thus dramatically increased the
size of the matrix which needs to be inverted. 
Fortunately, the range of the order parameter 
in $|{\bf x}-{\bf x'}|$ is truncated by 
the range of the pairing 
interaction.  For the isotropic-gap
superconductor we have 
}{
\rgtlne

\noindent
considered for most of this paper, the
effective pairing 
interaction is modeled by a delta function in 
space. It is possible to obtain 
anisotropic order parameters, including the $d$-wave
order parameters possibly appropriate for high-temperature
superconductors, merely by considering pairing with nearest neighbors
on a square tight-binding lattice. This model has been implemented with
the BdG formalism for 
non-magnetic impurities\cite{Xiang,Franz,Onishi,Flatunp} 
and 
magnetic impurities\cite{SalkBalat} and
vortices\cite{Soininen} in $d$-wave 
superconductors.  When
formulated on a lattice the addition of nearest-neighbor pairing
multiplies the rank of the matrix 
${\bf M^{n\rightarrow n}(\omega)}$ by $1+z$, 
where $z$ is the coordination
number of the lattice.
}

\twomini{

The order parameter's frequency
dependence complicates the self-consistency equation (Eq.~(\ref{KSsc})).
It must now be solved for each
frequency:
\begin{eqnarray}
\Delta&&({\bf r},{\bf r'};\omega)Z({\bf r},{\bf r'};\omega) \nonumber\\
&&\ \ = \int_{-\infty}^\infty d\epsilon n(\epsilon)\left[{\rm Re}F({\bf r},{\bf
r'};\epsilon)\right]\left(K_+({\bf r},{\bf r'};
\epsilon,\omega)-U_c\right)\nonumber\\
\omega&&\left[1-Z({\bf r},{\bf r'};\omega)\right] \nonumber\\&&\
\ =
\int_{-\infty}^\infty d\epsilon n(\epsilon)\left[{\rm Re}G({\bf r},{\bf
r'};\epsilon)\right]K_-({\bf r},{\bf r'};
\epsilon,\omega),\label{selflat}\end{eqnarray}
where $K_\pm$ are kernals of the pairing interaction and are
different for each mechanism of superconductivity.
They can be determined from the homogeneous solution.
 $U_c$ is a
Coulomb factor. $Z$, the quasiparticle weight, is solved
for self-consistently.
Incorporating these strong coupling effects allows a
determination of the 
effect of the frequency-dependence of the pairing
interaction on the electronic structure around a
defect. While selecting the particular location where
the STM tip is similar to selecting the momentum of the
quasiparticles of interest, selecting the STM voltage
indicates which order parameter frequency one wishes to probe.

\section{Summary}

The local electronic structure around an defect reflects
both the properties of the defect and the medium it is
embedded in.  For a strong non-magnetic or magnetic 
impurity in a superconductor,
the distortion of the
normal-state properties by the strong impurity plays a vital role
in the response of the superconducting medium. The LDOS for the
inhomogeneous superconductor can be related to the LDOS for the
inhomogeneous normal metal via equations like Eq.~(\ref{nonsum}) and
Eq.~(\ref{magsum}). In the case of a non-magnetic impurity with no localized
states around it, the LDOS for the inhomogeneous superconductor is
merely the normalized LDOS for the inhomogeneous normal metal multiplied by
the homogeneous superconductor's density of states. This should 
suggest some caution
for the method used to normalize STM spectra taken at different places on
a superconductor's surface.

For the case of a localized
state (such as around a magnetic impurity) with angular momentum
$\ell$, the LDOS for the state is the
angular-momentum-$\ell$-projected 
LDOS of 
\lftlne }{

\noindent
the inhomogeneous normal metal multiplied by a decaying exponential
whose range is determined by the energy of the localized state. Since the
spin-up band LDOS in 
the normal state differs from the spin-down 
band LDOS 
in the 
normal state, the electron-like pole of the localized quasiparticle will 
have 
different
spatial structure than the hole-like pole of the localized quasiparticle.

The self-consistent calculations described here have been performed with a 
new, powerful, Koster-Slater technique which allows the Gor'kov equation to
be solved in principle exactly. Although we have only presented calculations
for weak-coupling isotropic order parameters within a free-electron model,
we have formulated the extension of the new technique to strong-coupling 
pairing and general band structures and order parameter
symmetries. 

\section*{Acknowledgments}

We wish to thank C.M. Lieber for helpful conversations.
M.E.F. wishes to acknowledge the Office of Naval Research's
Grant No. N00014-96-1-1012. J.M.B. wishes 
to acknowledge an N.R.C. Postdoctoral Fellowship.
\vskip0.3pc

\section*{Appendix}

The expansion of the Green's functions of Eq.~(\ref{bareg})
suitable for a three-dimensional spherically symmetric
situation are detailed here. The homogeneous Green's
functions depend on ${\bf r}$ and ${\bf r'}$ through $r$,
$r'$, and $\cos\gamma = ({\bf r}\cdot{\bf r'})/rr'$. Then
the Green's functions can be expanded in Legendre
polynominals $P_\ell(\cos\gamma)$.
\begin{eqnarray}
g({\bf r},{\bf r'};\omega)  &&=\nonumber\\
g(r,r',\cos\gamma;\omega)
&&= {2\ell+1\over 4\pi}\sum_\ell
g_\ell(r,r';\omega)P_\ell(\cos\gamma)
\end{eqnarray}
and
\begin{equation}
g_\ell(r,r';\omega) = 2\pi\int_{-1}^1 P_\ell(x)g(r,r',x;\omega).\label{exp}
\end{equation}
Evaluating Eq.~(\ref{exp}) for both the normal and anomalous
Green's functions of Eq.~(\ref{bareg}) yields
}

\begin{eqnarray}
g_\ell(r,r';\omega) =&& -{\pi^3\over \sqrt{rr'}}\bigg[
i\left(1+{
\omega\over\sqrt{\omega^2-1}}\right)
J_{\ell+{1\over 2}}(\{1+\sqrt{\omega^2-1}/\xi\}r^<)
H^{(1)}_{\ell+{1\over 2}}(\{1+\sqrt{\omega^2-1}/\xi\}r^>)\nonumber\\
&& - \left(1-{
\omega\over\sqrt{\omega^2-1}}\right)
J_{\ell+{1\over 2}}(\{1-\sqrt{\omega^2-1}/\xi\}r^<)
H^{(2)}_{\ell+{1\over 2}}(\{1-\sqrt{\omega^2-1}/\xi\}r^>)\bigg]\\
f_\ell(r,r';\omega) =&& -{i\pi^3\over \sqrt{rr'}}{1\over\sqrt{
\omega^2-1}}\bigg[
J_{\ell+{1\over 2}}(\{1+\sqrt{\omega^2-1}/\xi\}r^<)
H^{(1)}_{\ell+{1\over 2}}(\{1+\sqrt{\omega^2-1}/\xi\}r^>)\nonumber\\
&& + 
J_{\ell+{1\over 2}}(\{1-\sqrt{\omega^2-1}/\xi\}r^<)
H^{(2)}_{\ell+{1\over 2}}(\{1-\sqrt{\omega^2-1}/\xi\}r^>)\bigg]
\end{eqnarray}
\twomini{\vskip0.3pc
\noindent where $J_\ell$, $H^{(1)}_\ell$, and $H^{(2)}_\ell$ are standard
Bessel functions, $r^<$ ($r^>$) is the smaller (larger) of $r$ and $r'$,
$\omega$ is in units of $\Delta_o$ and $r$ and $r'$ are in
units of $k_F^{-1}$. The Green's functions are in units of
$N_o$.

The Gor'kov equation, Eq.~(\ref{KS}), can now be written in a
form diagonal in $\ell$,
\begin{eqnarray}
{\bf G}_\ell&&(r,r';\omega) = \\&&{\bf g}_\ell(r,r';\omega) 
+ \int_0^\infty r_n^2dr_n
{\bf g}_\ell(r,r_n;\omega) {\bf V}(r_n)
{\bf G}_\ell(r_n,r';\omega).\nonumber
\end{eqnarray}
}{\vskip0.5pc

\noindent
Thus the three-dimensional integral has been reduced to a one-dimensional 
radial integral.
Since the numerical inversion procedure depends on inverting a matrix 
whose rank is proportional
to the number of spatial
points considered, this reduction to a one-dimensional integral
dramatically increases the speed of this calculation over
a calculation for a three-dimensional potential which is not spherically 
symmetric.
}


\begin{references}
\bibitem{noteus} When the calculation is performed self-consistently,
as will be described later in the paper, a very weakly-localized
state may exist around a non-magnetic impurity.
\bibitem{topol} Topological defects, such as vortices, can
be different.
\bibitem{FMAFM} Unlike the case of a quantum spin, there is
no qualitative difference between ferromagnetic and
antiferromagnetic coupling for the classical spin. We
therefore do not discuss the orientation of the impurity
spin, but merely define a low-energy direction for the
quasiparticle spin.
\bibitem{usefulQ} This might be useful if there is a
technical problem with making measurements in the normal
state.
\end{references}
\end{document}